\documentclass[pra,11pt,twoside]{revtex4}
\usepackage{amssymb}
\usepackage{graphics}
\usepackage{ifthen}
\renewcommand{\baselinestretch}{1.1}
\begin{document}

\title{Nonlinear Wightman fields}
\author{Peter Morgan}
\email{peter.w.morgan@yale.edu}
\affiliation{Physics Department, Yale University, New Haven, CT 06520, USA.}
\homepage{http://pantheon.yale.edu/~PWM22}

\date{\today}
\begin{abstract}
A nonlinear Wightman field is taken to be a nonlinear map from a linear space of test functions to a linear space of
Hilbert space operators, with inessential modifications to other axioms only to the extent dictated by the introduction
of nonlinearity.
Two approaches to nonlinear quantum fields are constructed and discussed, the first of which, starting from
Lagrangian QFT, offers a fresh perspective on renormalization, while the second, starting from linear Wightman
fields, provides an extensive range of well-defined nonlinear theories.
\end{abstract}

\maketitle

\newcommand\Half{{\frac{1}{2}}}
\newcommand\Intd{{\mathrm{d}}}
\newcommand\Remove[1]{{\raise 1.2ex\hbox{$\times$}\kern-0.8em \lower 0.35ex\hbox{$#1$}}}
\newcommand\SmallFrac[2]{{\scriptstyle\frac{\scriptstyle #1}{\scriptstyle#2}}}
\newcommand\BLow[1]{{\lower 0.65ex\hbox{${}_{#1}\!\!$}}}
\newcommand\Vacuum{{\left|0\right>}}
\newcommand\VEV[1]{{\left<0\right|#1\left|0\right>}}
\newcommand\iD{{\mathrm{i}\hspace{-1.5pt}\Delta}}
\newcommand\itD{{\mathrm{i}\hspace{-1.5pt}\tilde\Delta}}
\newcommand\xxi{{\hat\xi\hspace{-0.85ex}\xi}}
\newcommand\zzeta{{\hat\zeta\hspace{-0.85ex}\zeta}}
\newcommand\subxxi{{\xi\hspace{-0.65ex}\xi}}
\newcommand\subzzeta{{\zeta\hspace{-0.65ex}\zeta}}
\newcommand\NewSection{{}}
\newcommand\nlIP[1]{{(\!(#1)\!)}}
\newcommand\Lbbeta{{\mbox{\boldmath$\scriptstyle\beta$\unboldmath}}}
\newcommand\bbeta{{\mbox{\boldmath$\scriptstyle\beta$\unboldmath}}}
\newcommand\sbbeta{{\mbox{\boldmath$\scriptscriptstyle\beta$\unboldmath}}}
\newcommand\eqdef{{\stackrel{\scriptstyle\mathrm{def}}{=}}}

\section{Introduction}
The reconciliation of the mathematically coherent Wightman axioms with the empirically successful Lagrangian and
related approaches to interacting quantum field theory is a longstanding problem~\cite{Fredenhagen2007,Buchholz2000}.
The nature of the mismatch between interacting quantum fields such as QED and the Standard Model of particle
physics and the Wightman axioms is not clear, however it appears that the introduction of products of
distributions in Lagrangian QFT is one heart of the difficulty.
Because Lagrangian QFT introduces nonlinear interaction terms into the Lagrangian without changing the linearity
of the underlying Hilbert space, we are motivated to consider what similar forms of nonlinearity might be
introduced into the Wightman or Haag-Kastler axioms.
Real-space renormalization also motivates a nonlinear structure because processes that are used to construct
higher-level block variables, such as \emph{majority rule} and \emph{decimation}, are not in general expressible
as weighted averages of lower-level block variables.

The postulate that a Wightman field $\hat\phi(x)$ is an operator-valued \emph{distribution}, so that the operators
of the theory are obtained by linear ``smearing'', for a well-behaved linear space $\mathcal{S}$ of ``test'' functions,
\begin{equation}\label{QFdefinition}
  \hat\phi:\mathcal{S}\rightarrow\mathcal{A};f\mapsto\hat\phi_f= \int\hat\phi(x)f(x)\Intd^4x,
\end{equation}
has hitherto not been questioned in the literature.
Our point of departure will be to take $\hat\xi:f\mapsto\hat\xi_f$, for a suitably well-behaved linear space of
test functions, to be a nonlinear functional of the test functions, so that in general
$\hat\xi_{f+g}\not=\hat\xi_f+\hat\xi_g$ and $\hat\xi_{\lambda f}\not=\lambda\hat\xi_f$, essentially denying the
linearity that is implied by Eq. (\ref{QFdefinition}).
Nonetheless, the action of the $*$-algebra of operators on Hilbert space vectors is taken to be linear,
$(\hat\xi_f+\hat\xi_g)\left|\psi\right>=\hat\xi_f\left|\psi\right>+\hat\xi_g\left|\psi\right>$ and
$(\lambda\hat\xi_f)\left|\psi\right>=\hat\xi_f(\lambda\left|\psi\right>)$, and we will retain the usual Born rule
construction of expected values, of probabilities, and of correlations that is common to all quantum theory ---and their
use to model the statistics of multiple experimental datasets--- and the converse GNS-construction of a Hilbert space from
the expected values that are generated by a state over a $*$-algebra of operators~\cite[\S III.2]{Haag}.
This ``Born-GNS'' linearity is necessary, as in Lagrangian QFT, so that quantum theory generates probabilities that satisfy the
Kolmogorov axioms of probability, however the linearity we relinquish has no comparable necessity.
In the construction introduced in Section \ref{CAalgebra}, we will be able to preserve the conditions that are satisfied
by a Wightman field ---Cluster Decomposition, Relativistic Transformation, Spectrum, Hermiticity, Local Commutativity,
and Positive Definiteness---, and that are required of the Vacuum Expectation Values (VEVs) to allow the use of the
Wightman reconstruction theorem to construct a Wightman field~\cite[\S 3-4]{SW}, in forms that are appropriate
for nonlinear Wightman fields.

It is of course possible to question other postulates.
Numerous ways to break the Lorentz invariance of the dynamics, both at very large distances and at very short
distances, have been proposed, however we here suppose the dynamics to be Lorentz invariant, on a Minkowski space background,
ruling out for our purposes, in particular, the concerns of quantum field theory in curved space-time and of quantum gravity.
Streater~\cite[\S 3.4]{Streater1975} discusses a number of ways in which the Wightman axioms have been questioned, but
there is no introduction of nonlinearity of the form proposed here, nor either is there in the more recent
discussions mentioned above~\cite{Fredenhagen2007,Buchholz2000}.
The equivalent Haag-Kastler axiom is the additivity property, which requires that the operator algebra associated
with two regions of space-time is generated by the operator algebras associated with the two independent
regions~\cite[\S III.1]{Haag}, however we will work here more concretely with (nonlinear) Wightman fields.
From the direction of Lagrangian QFT, the mathematics has somewhat improved with the introduction of a Hopf algebraic
approach to Feynman diagrams~\cite{EFK}, however, again, we prefer here to begin with Wightman fields.

We will introduce two very different nonlinear constructions, one, non-rigorous, that approximately follows and is motivated
by the method that is used by Lagrangian QFT to construct an interacting quantum field, in Section \ref{NonlinearQFT}, and
another, in Section \ref{CAalgebra}, that modifies the free field $*$-algebra of creation and annihilation operators.
Section \ref{SomeStandardQFT} describes some standard interacting QFT, then presents the same material in a way
that motivates the construction in Section \ref{NonlinearQFT}.

\NewSection
\section{Some standard Quantum Field Theory}\label{SomeStandardQFT}
\renewcommand{\baselinestretch}{1.15}\normalsize
The direct textbook way to construct an interacting scalar quantum field is to introduce a formal time-dependent
transformation of a free quantum field $\hat\phi(x)$~\cite[\S 6-1-1]{IZ},
\begin{eqnarray}\label{XiDefinition}
  \hat\xi(x)&=&\hat U^{-1}(x_0)\hat\phi(x)\hat U(x_0)\hspace{6em}\mathrm{where}\ 
      \hat U(\tau)=\mathsf{T}\!\left[\!\!\raisebox{-0.9ex}{
                       $\mathrm{e}^{-\mathrm{i}\!\!\!\int\limits_{-\infty}^\tau\!\!\!\hat H(y)\Intd^4y}$}\right],\cr
            &=&\mathsf{T}^\dagger\!\left[\!\!\raisebox{-0.9ex}{
                       $\mathrm{e}^{-\mathrm{i}\!\!\!\int\limits_{-\infty}^\infty\!\!\!\hat H(y)\Intd^4y}$}\right]
               \mathsf{T}\!\left[\!\!\raisebox{-0.9ex}{
                       $\hat\phi(x)\mathrm{e}^{-\mathrm{i}\!\!\!\int\limits_{-\infty}^\infty\!\!\!\hat H(y)\Intd^4y}$}\right],
\end{eqnarray}
where $\hat H(y)$ is a local operator, constructed as a sum of normal-ordered products of $\hat\phi(y)$ and its derivatives.

A standard first step is a heuristic derivation of the time-ordered VEVs for the interacting field,
\begin{equation}\label{TimeOrderedVEVs}
  \VEV{\mathsf{T}\!\left[\hat\xi(x_1)\cdots\hat\xi(x_n)\right]}=
       \frac{\VEV{\mathsf{T}\!\left[\!\!\raisebox{-0.9ex}{
                       $\hat\phi(x_1)\cdots\hat\phi(x_n)
                                  \mathrm{e}^{-\mathrm{i}\!\!\!\int\limits_{-\infty}^\infty\!\!\!\hat H(y)\Intd^4y}$}\right]}}
            {\VEV{\mathsf{T}\!\left[\!\!\raisebox{-0.9ex}{
                       $\mathrm{e}^{-\mathrm{i}\!\!\!\int\limits_{-\infty}^\infty\!\!\!\hat H(y)\Intd^4y}$}\right]}},
\end{equation}
which in turn allows the construction of Feynman integrals and the corresponding Feynman diagrams.
As well as the formal nature of this construction, also the time-ordering of the left-hand side of this equation
compromises its connection to the Wightman axioms, for which one foundation is a set of \emph{non}-time-ordered VEVs
that are distributions and that satisfy six sets of Conditions ---Cluster Decomposition, Relativistic Transformation,
Spectrum, Hermiticity, Local Commutativity, and Positive Definiteness---, which would allow the use of the Wightman
reconstruction theorem to construct a Wightman field.

A standard alternative is to construct an S-matrix, which time-evolves vector-valued distributions such as
$\hat\phi(x_1)\cdots\hat\phi(x_n)\Vacuum$ (or the equivalent in wave-number coordinates), which form an
improper basis for the free field Hilbert space of initial states, using the formal construction
$\mathsf{T}\!\left[\!\!\raisebox{-0.9ex}{
                       $\mathrm{e}^{-\mathrm{i}\!\!\!\int\limits_{-\infty}^\infty\!\!\!\hat H(y)\Intd^4y}$}\right]$
as the evolution operator (introducing an idealization of infinite time-separation between preparation and measurement
or between initial and final vector states that is reasonable in particle physics experiments but that is not generally
satisfied), then use the inner product with a final state, a vector in a Hilbert space that is taken to be unitarily
equivalent to the free field Hilbert space of initial states.
We obtain constructions such as
\begin{equation}\label{SmatrixExpression}
  \VEV{\hat\phi(y_m)\cdots\hat\phi(y_1)
         \mathsf{T}\!\left[\!\!\raisebox{-0.9ex}{
                       $\mathrm{e}^{-\mathrm{i}\!\!\!\int\limits_{-\infty}^\infty\!\!\!\hat H(y)\Intd^4y}$}\right]
       \hat\phi(x_1)\cdots\hat\phi(x_n)},
\end{equation}
which allow us to compute transition probabilities.
This, however, also cannot be used to construct a Wightman field.

Neither the time-ordered VEVs construction nor the S-matrix construction is well-defined, because, at least,
the integral ${\scriptstyle \int\limits_{-\infty}^\infty\!\!\!\hat H(y)\Intd^4y}$ does not exist, however regularization and
renormalization allows these and other constructions to make contact with experiment, which is unsurprisingly
taken by most physicists to be more important than making contact with the Wightman axioms.

\subsection{A different construction}\label{DifferentQFT}
Here we directly construct non-time-ordered VEVs for the interacting quantum field by noting that the
components of $\hat U(x_0)$ that are space-like separated from $x$ commute with $\hat\phi(x)$, and because of time-ordering
those components cancel with the time-reversed components of the inverse $\hat U^{-1}(x_0)$, so that the
interacting field $\hat\xi(x)$ can also be written Lorentz invariantly as
\begin{equation}\label{xiDefinition}
  \hat\xi(x)=\mathsf{T}^\dagger\!\left[\mathrm{e}^{-\mathrm{i}\hat\mathcal{L}(x)}\right]\hat\phi(x)
                          \mathsf{T}\!\left[\mathrm{e}^{-\mathrm{i}\hat\mathcal{L}(x)}\right],\qquad
  \mathrm{where}\ \hat\mathcal{L}(x)= \int\limits_{\blacktriangle(x)}\!\hat H(y)\Intd^4y
\end{equation}
and $\blacktriangle(x)=\{y:(x-y)^2\ge 0\ \mathrm{and}\ x_0>y_0\}$ is the region of space-time that is light-like or
time-like separated from and earlier than $x$.
Furthermore, the adjoint action of $\hat\phi(x)$ on a time-ordered expression is a derivation, because time-ordering
ensures commutativity, so that, taking a quartic scalar interaction with Hamiltonian density
$\SmallFrac{\lambda}{4!}\!:\!\hat\phi^4(y)\!:$ as an example,
$$
  \left[\hat\phi(x),\mathsf{T}\!\left[\left(\int\!:\!\hat\phi^4(y)\!:\Intd^4y\right)^n\right]\right]=
       \mathsf{T}\!\left[\int\!4n\iD(x-z):\!\hat\phi^3(z)\!:\Intd^4z\left(\int\!:\!\hat\phi^4(y)\!:\Intd^4y\right)^{n-1}\right],
$$
where
\begin{equation}\label{iDG}
  \iD(x-z)=-\mathrm{i}(G_{\mathrm{ret}}(x-z)-G_{\mathrm{adv}}(x-z))=[\hat\phi(x),\hat\phi(z)]
\end{equation}
is the free field commutator and $G_{\mathrm{ret}}(x-z)$ and $G_{\mathrm{adv}}(x-z)$ are the retarded and advanced Green
functions~\cite[\S 1-3-1]{IZ}.
For the interacting field, therefore, we have the construction
\begin{eqnarray}
  \hat\xi(x)&=&\mathsf{T}^\dagger\!\left[\mathrm{e}^{-\mathrm{i}\hat\mathcal{L}(x)}\right]\hat\phi(x)
                          \mathsf{T}\!\left[\mathrm{e}^{-\mathrm{i}\hat\mathcal{L}(x)}\right]\cr
            &&\hspace{3em}\mbox{[$\hat\phi(x)$ acts as a derivation, ...]}\cr
            &=&\mathsf{T}^\dagger\!\left[\mathrm{e}^{-\mathrm{i}\hat\mathcal{L}(x)}\right]
                       \left(\!\mathsf{T}\left[\mathrm{e}^{-\mathrm{i}\hat\mathcal{L}(x)}\right]\hat\phi(x)
                          -\mathsf{T}\!\left[\SmallFrac{\mathrm{i\lambda}}{3!}\!\!\!
                              \int\limits_{\blacktriangle(x)}\!\iD(x-z):\!\hat\phi^3(z)\!:\Intd^4z
                                    \mathrm{e}^{-\mathrm{i}\hat\mathcal{L}(x)}\right]\right)\cr
            &=&\hat\phi(x)-\mathsf{T}^\dagger\!\left[\mathrm{e}^{-\mathrm{i}\hat\mathcal{L}(x)}\right]
                          \mathsf{T}\!\left[\SmallFrac{\mathrm{i\lambda}}{3!}\!\!\!
                              \int\limits_{\blacktriangle(x)}\!\iD(x-z):\!\hat\phi^3(z)\!:\Intd^4z
                                    \mathrm{e}^{-\mathrm{i}\hat\mathcal{L}(x)}\right]\cr
            &=&\hat\phi(x)-\SmallFrac{\mathrm{i\lambda}}{3!}\!\!\!\int\limits_{\blacktriangle(x)}\iD(x-z)
                                \mathsf{T}^\dagger\!\left[\mathrm{e}^{-\mathrm{i}\hat\mathcal{L}(x)}\right]
                                \mathsf{T}\!\left[:\!\hat\phi^3(z)\!:\mathrm{e}^{-\mathrm{i}\hat\mathcal{L}(x)}\right]\Intd^4z\cr
            &&\hspace{3em}\mbox{[components of $\hat\mathcal{L}(x)$ that are space-like separated from $z$}\cr
            &&\hspace{13em}\mbox{or are later than $z$ \emph{cancel}, leaving $\hat\mathcal{L}(z)$, ...]}\cr
            &=&\hat\phi(x)-\SmallFrac{\mathrm{i\lambda}}{3!}\!\!\!\int\limits_{\blacktriangle(x)}\iD(x-z)
                                \mathsf{T}^\dagger\!\left[\mathrm{e}^{-\mathrm{i}\hat\mathcal{L}(z)}\right]
                                     :\!\hat\phi^3(z)\!:
                                \mathsf{T}\!\left[\mathrm{e}^{-\mathrm{i}\hat\mathcal{L}(z)}\right]\Intd^4z\cr
            &=&\hat\phi(x)-\SmallFrac{\mathrm{i\lambda}}{3!}\!\!\!
                              \int\limits_{\blacktriangle(x)}\!\iD(x-z):\!\hat\xi^3(z)\!:\Intd^4z\quad
        \biggl[:\!\hat\xi^3(z)\!: \;=\mathsf{T}^\dagger\!\left[\mathrm{e}^{-\mathrm{i}\hat\mathcal{L}(z)}\right]
                                     :\!\hat\phi^3(z)\!:
                                \mathsf{T}\!\left[\mathrm{e}^{-\mathrm{i}\hat\mathcal{L}(z)}\right]\biggr]\cr
            &=&\hat\phi(x)-\SmallFrac{\lambda}{3!}\!\!
                              \int\!G_{\mathrm{ret}}(x-z):\!\hat\xi^3(z)\!:\Intd^4z,\label{IntegralEqn}
\end{eqnarray}
where the restriction to the backward light-cone $\blacktriangle(x)$ is equivalent to replacing the propagator
$\iD(x-z)$ by $-\mathrm{i}G_{\mathrm{ret}}(x-z)$, as we see from Eq. (\ref{iDG}).
Insofar as we can take $:\!\hat\phi^3(z)\!:$ formally to be an infinite multiple of $\hat\phi(z)$ subtracted
from $\hat\phi^3(z)$, we can take $:\!\hat\xi^3(z)\!:$ formally to be the same infinite multiple of $\hat\xi(z)$
subtracted from $\hat\xi^3(z)$.
$\hat\phi(x)$ satisfies the homogeneous Klein-Gordon equation, $(\Box+m^2)\hat\phi(x)=0$, and
$G_{\mathrm{ret}}(x-z)$ satisfies $(\Box+m^2)G_{\mathrm{ret}}(x-z)=\delta^4(x-z)$, so, applying the operator
$(\Box+m^2)$ to the last line of Eq. (\ref{IntegralEqn}), $\hat\xi(x)$ satisfies the
nonlinear differential equation
\begin{eqnarray}\label{FieldDiffEqn}
  (\Box+m^2)\hat\xi(x)+\SmallFrac{\lambda}{3!}:\!\hat\xi^3(x)\!:\;=0.
\end{eqnarray}
The above construction shows that, apart from the mathematical necessity to regularize and renormalize, to
construct an interacting field we may replace $\hat\phi(x)$ at a point by a complex of operators at
points of the backward light-cone of $x$, constructed using the propagator $G_{\mathrm{ret}}(x-z)$.

Once we have constructed $\hat\xi(x)$ as a functional of $\hat\phi(z)$, using $G_{\mathrm{ret}}(x-z)$ to propagate
to all points in the backward light-cone, we may use Wick's theorem~\cite[\S 4-2]{IZ} formally and the positive
frequency propagator $\itD_+(k)=2\pi\delta(k\!\cdot\!k-m^2)\theta(k_0)$ to put the complex of operators that is associated
with $\blacktriangle(x)$ into normal-ordered form,
\begin{equation}\label{xiNO}
  \hat\xi(x)=\sum\limits_{n=1}^\infty\int G_n(x-y_1,...,x-y_n):\!\hat\phi(y_1)\cdots\hat\phi(y_n)\!:
                                         \Intd^4y_1\cdots\Intd^4y_n,
\end{equation}
where the Lorentz covariant functions $G_n(\cdot)$, which may be taken to be symmetric in their arguments and
after renormalization must be such that VEVs are finite, are determined by whatever Hamiltonian we use for the
interacting theory, and are zero unless every $x-y_i$ that is a parameter of $G_n(x-y_1,...,x-y_n)$ is in or on
the forward light-cone.
Having constructed the operator $\hat\xi(x)$ in normal-ordered form, the formal Wick contraction of a product
$\hat\xi(x_1)\hat\xi(x_2)$ of two such operators yields a normal-ordered operator that is a complex of operators
associated with the backward light-cones both of $x_1$ and of $x_2$, which in principle determines the commutation
relations for the interacting field, again, however, up to regularization and renormalization.
To construct $n$-point VEVs, we again may use Wick's theorem and the positive frequency propagator, now between the
infinite complexes of operators that are associated with the $n$ backward light-cones of the original $n$ points.
This contrasts with the time-ordered VEVs of Eq. (\ref{TimeOrderedVEVs}), which use only the Feynman propagator;
and with the S-matrix construction of Eq. (\ref{SmatrixExpression}), which uses the positive frequency propagator
whenever one of a pair of operators is associated with the initial or final state vectors, because the initial and final
state vectors are not within the scope of the time-ordering, and uses the Feynman propagator between all pairs of points
that occur within the time-ordering.
Nonetheless, all three constructions are quite direct consequences of Eq. (\ref{XiDefinition}).

\begin{figure}[tb]
  \includegraphics{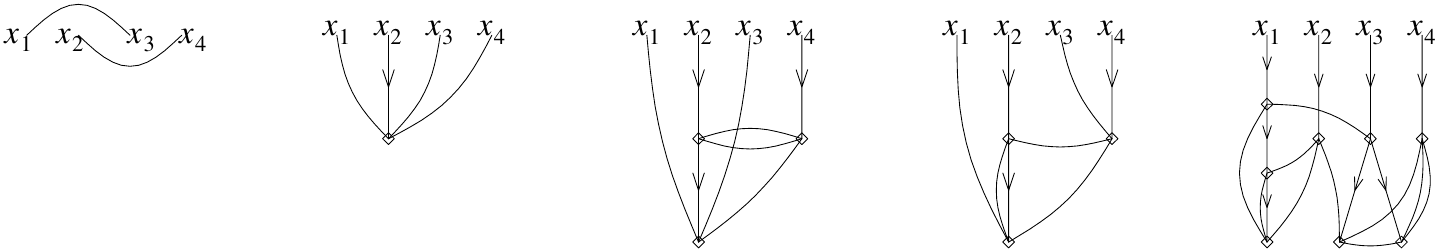}
\caption{\label{Mgraphs2}Some example 4-point graphs that arise from $\VEV{\hat\xi(x_1)\hat\xi(x_2)\hat\xi(x_3)\hat\xi(x_4)}$,
depicting propagators $G_{\mathrm{ret}}$ using arrows, which are non-zero only at light-like or time-like separation, and
propagators $\iD_+$, using curved lines, which are non-zero for any space-time separation.
Time increases from bottom to top, but the points $x_1$, $x_2$, $x_3$, $x_4$, and any points that are not joined by arrows,
may be at arbitrary mutual space-time separation.}
\end{figure}

\begin{figure}[tb]
  \includegraphics{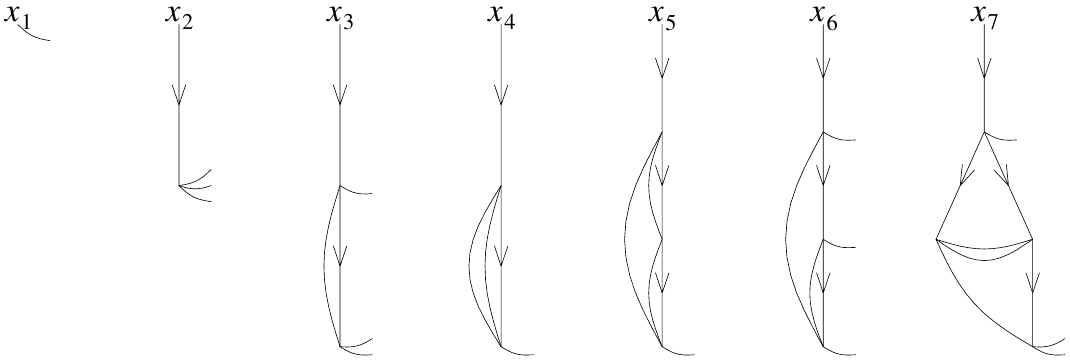}
\caption{\label{Mgraphs1}Some example graphs that contribute to expansions of $\hat\xi(x_1)$, ... .
The short unconnected edges must all connect with unconnected edges from other such expansions when computing contributions to VEVs.}
\end{figure}

The distinction between the two propagators $\iD_+(x)$ and $G_{\mathrm{ret}}(x)$, the first being confined to
the forward light-cone in wave-number space and the second being confined to the forward light-cone in real space,
gives an alternative to conventional discussions of virtual particles, in that we have used the
virtual propagator $G_{\mathrm{ret}}(x)$ to construct an interacting field operator as a complex of free field
operators that are confined to the backward light-cone.
The form of Eq. (\ref{IntegralEqn}) also suggests a form of diagram different from that of Feynman diagrams, in
which we distinguish between the propagators $\iD_+(x)$ and $G_{\mathrm{ret}}(x)$ in the integrals that arise.
Some examples are given in Fig. \ref{Mgraphs2}.
We may also construct graphs, as in Fig. \ref{Mgraphs1}, that contribute to the terms that arise in Eq. (\ref{xiNO}).
This alternative discussion does not undermine conventional discussions or computations, but it provides an
alternative that takes an interacting field operator $\hat\xi(x)$ to be associated with free field operators in
the backward light-cone of $x$, instead of taking a particle to be surrounded by a sea of virtual particles at
the same time.
This approach is possible insofar as we take the structure of quantum field theory to be a single Hilbert space that is
associated with 4-dimensional space-time, which is standard for Wightman fields, instead of the alternative, in which
we take the structure to be a set of Hilbert spaces that are individually associated with a phase space at a given time,
with unitary evolution acting on a given Hilbert space to produce Hilbert spaces associated with different times.
Although standard discussions of Lagrangian QFT begin with Hilbert spaces associated with phase spaces, calculations
in Lagrangian QFT are compatible with either structure.

It is essential that time-ordering ensures that the complexes of operators in the backward light-cones of two
space-like separated points do not modify the commutativity of the interacting field at space-like separation, because
the product $\hat\xi(x)\hat\xi(y)$ can be written as 
\begin{eqnarray}\label{Unitary}
  \hat U^{-1}(x_0)\hat\phi(x)\hat U(x_0)\hat U^{-1}(y_0)\hat\phi(y)\hat U(y_0)&=&
      \hat U^{-1}(y_0)\hat U(y_0)\hat U^{-1}(x_0)\hat\phi(x)\hat U(x_0)\hat U^{-1}(y_0)\hat\phi(y)\hat U(y_0)\cr
   &=&\hat U^{-1}(y_0)\hat U^{-1}(x_0,y_0)\hat\phi(x)\hat U(x_0,y_0)\hat\phi(y)\hat U(y_0).
\end{eqnarray}
When $x$ and $y$ are space-like separated, we can note that there is a choice of time coordinates in which $x_0=y_0$,
$\hat U(x_0,y_0)= \hat U(x_0)\hat U^{-1}(y_0)=1$, or we can, alternatively, note that for arbitrary time coordinates
$\hat U^{-1}(x_0,y_0)\hat\phi(x)\hat U(x_0,y_0)$ replaces $\hat\phi(x)$ by a complex of free field operators at points
of the backward light-cone at $x$ that are all at space-like separation from $y$.
Hence, as for the free field $\hat\phi(x)$, $\hat\xi(x)$ satisfies $[\hat\xi(x),\hat\xi(y)]=0$ when $x$ and $y$ are
space-like separated.
It is taken to be a fundamental constraint on renormalization that commutativity of the renormalized field at
space-like separation is preserved.

Regularization and renormalization of this construction is as much a problem as it is for the conventional
construction of time-ordered VEVs and of the S-matrix, however what we are constructing is somewhat clearer,
and closer to the Wightman axioms in character, because we have reduced some of the distraction introduced by
time-ordering.

\NewSection
\section{A first nonlinear construction}\label{NonlinearQFT}
As well as ``smearing'' with a test function, we may also take some motivation from signal analysis to construct
a convolution of a quantum field with a smooth window function $w(\cdot)$,
\begin{equation}
  \hat\phi_w(x)= [\hat\phi\star w](x)=\int\hat\phi(x-y)w(y)\Intd^4y=\int\hat\phi(y)w(x-y)\Intd^4y,
\end{equation}
which for the free field satisfies the linear differential equation $(\Box+m^2)\hat\phi_w(x)=0$ for any window function
and for which we have the commutation relations $[\hat\phi_{w_1}(x),\hat\phi_{w_2}(y)]=[w_1\star\iD\star w_2^{(-)}](x-y)$,
where $w^{(-)}(z)= w(-z)$.
We could equally well present this convolution in terms of smearing as $\hat\phi_w(x)=\hat\phi_{w_x}$,
where we define the indexed set of test functions $w_x$ as $w_x(y)= w(x-y)$, however particle physics
has become something of an extreme exercise in processing large numbers of electronic signals, where window
functions are at least as familiar as test functions.

The process of smearing and of convolution is fundamental to the renormalization group, as described by Wilson and Kogut,
for example,
\begin{quote}``The [first] basic idea [of the renormalization group] is the same as in hydrodynamics.
In hydrodynamics one introduces new variables such as the density $\rho(x)$ which represents an average over the original
microscopic degrees of freedom.''~\cite[p. 79]{WilsonKogut}
\end{quote}
It is supposed here that such a smeared classical density $\rho(x)$ ought to be indexed by a description of the averaging
process, for which in the linear case we might choose a window function $w$, giving us $\rho_w(x)$.
The process of real-space renormalization, furthermore, introduces a hierarchy of smeared observables, in which successive
levels are constructed from the level below, and operators associated with successive levels of the hierarchy are taken to
satisfy different equations.
The practice is to construct successive levels, typically nonlinearly, by applying a single process repeatedly, so that
successive smeared observables may be indexed by a single scale, however each level may in principle be an arbitrary
nonlinear functional of the levels below.
We here take an operator $\hat\xi_w(x)$ to be a nonlinear functional of $w$, but we will discuss only marginally
how $\hat\xi_w(x)$ might be obtained from other $\hat\xi_{w'}(x)$, ..., instead focusing abstractly on the VEVs that are
obtained for a given window function.

A deformation that follows the construction of Section \ref{DifferentQFT} requires an interacting quantum field
$\hat\xi_w(x)$ to satisfy a nonlinear differential equation that depends on the window function, such as, for
the $\phi^4$ potential,
\begin{equation}\label{windowEq}
  (\Box+m^2)\hat\xi_w(x)+\SmallFrac{\lambda[w]}{3!}\left(\hat\xi_w(x)^3-3\mu[w]\hat\xi_w(x)\right)=0,
\end{equation}
where the interaction functionals $\lambda[w]$ and $\mu[w]$ should be manifestly Poincar\'e invariant and the term
$-3\mu[w]\hat\xi_w(x)$ has a similar effect to that of normal-ordering.
Given this covariant construction, which accords with the usual idea that the Lagrangian changes as we consider
different renormalization scales, $\hat\xi_w(x)$ in general cannot be a linear functional of $w_x$.
Note that a governing idea of effective field theory is that in principle all possible interaction terms that are consistent
with experimentally observed symmetries should be present in such a differential equation for an effective observable, not
just the two interaction terms given in Eq. (\ref{windowEq}), and that all the interaction strengths should be functions
of the details of the construction of the effective observables~\cite[\S 8 and refs. therein]{Wells}, so that for an
effective field theory we might work with
\begin{equation}
  (\Box+m^2)\hat\xi_w(x)+\sum\limits_{j=1}^\infty\lambda_j[w]\hat\xi_w(x)^j=0,
\end{equation}
or, more generally, a functional $F_w$ of arbitrary Lorentz invariant combinations of derivatives of $\hat\xi_w(x)$ should
be introduced,
\begin{equation}\label{XiEquation}
  (\Box+m^2)\hat\xi_w(x)+F_w\left(\hat\xi_w(x),\frac{\partial\hat\xi_w(x)}{\partial x_\mu}\frac{\partial\hat\xi_w(x)}{\partial x^\mu},
      \frac{\partial^2\hat\xi_w(x)}{\partial x_\mu\partial x_\nu}\frac{\partial^2\hat\xi_w(x)}{\partial x^\mu\partial x^\nu},...\right)=0.
\end{equation}
Particular solutions of Eq. (\ref{windowEq}) may be constructed perturbatively as
\begin{equation}
            \hat\xi_w(z)=\hat\phi_w(x)-\SmallFrac{\lambda[w]}{3!}\!\!
                              \int\!G_{\mathrm{ret}}(x-z)\left(\hat\xi_w(z)^3-3\mu[w]\hat\xi_w(z)\right)\Intd^4z.
\end{equation}
When $w(\cdot)$ is not close to the Dirac $\delta$-function, there is no guarantee that $\hat\xi_w(x)$
commutes with $\hat\xi_w(y)$ when $x$ and $y$ are space-like separated enough to make the supports of $w_x$
and $w_y$ space-like separated, because we cannot use Eq. (\ref{Unitary}), but when $w(\cdot)$ approaches the
Dirac $\delta$-function, Eq. (\ref{windowEq}) approaches the conventional Eq. (\ref{FieldDiffEqn}), with
potentially infinite $\lambda[w]$ and $\mu[w]$, and $\hat\xi_w(x)$ approaches the $\hat\xi(x)$ of Eq. (\ref{xiDefinition}).
The functional dependence of $\lambda[w]$ and of other interaction constants can be taken to be an explicit way to control
the mathematics of regularization and renormalization, with the infinite-dimensional dependence on $w(\cdot)$ replacing
the usual dependence on the single-dimensional renormalization scale; different functional properties of $\lambda[w]$ can
be used to tune the variation of VEVs for different choices of $w$, and modification of the functional dependence of the
mass term $\mu[w]$ may also be useful.
In these terms, real-space renormalization determines an operator-valued function $\hat\xi_w$ as a functional of
$\hat\xi_{S_1[w]}$, $\hat\xi_{S_2[w]}$, ..., where $S_1[w]$, $S_2[w]$, ... are functionals of $w$; for example, but
with more general constructions possible,
\begin{equation}
  \hat\xi_w(x)=\hspace{-1.5em}\int\limits_{\mathrm{Supp}(S_i[w]_{z_i})\subset \mathrm{Supp}(w_x)\;\forall i}\hspace{-1.5em}
                  R_w\left(\hat\xi_{S_1[w]}(z_1), \hat\xi_{S_2[w]}(z_2), ...\right)\rho_{w_x}(z_1,z_2,...)\prod\limits_i\Intd^4z_i,
\end{equation}
which in general will not be reversible, in such a way that $\hat\xi_{w'}(x)$ satisfies a Poincar\'e covariant equation
of the same structure as $\hat\xi_w(x)$, but with (a preferably finite number of) coefficients $\lambda_j[w']$
replacing $\lambda_j[w]$.
One example of an increase in generality is for the integral to be restricted only to the causal completion,
$\mathrm{Supp}(S_i[w]_{z_i})\subset \mathrm{Supp}(w_x)''\supseteq \mathrm{Supp}(w_x)$.
The density $\rho_{w_x}(z_1,z_2,...)$ must be a translation and Lorentz covariant functional of $w_x$, which could be taken,
as for straightforward blocking, to be zero except for $z_i$ on a lattice of points, but we could also allow
$\rho_{w_x}(z_1,z_2,...)$ to be continuous; similarly, as effective models we might take $\hat\xi_w(x)$ to be defined only
for $x$ on a lattice of points without concern at the consequent lack of translation invariance, but we could also allow
$\hat\xi_w(x)$ to be defined for any $x$.
There is a tacit understanding that the momentum space forms of regularization and renormalization that are used in quantum
field theory are more-or-less \emph{sui generis} with real-space renormalization, however an attempt to place subtle analytic
approaches such as dimensional regularization accurately in such a relationship is taken to be beyond the scope of this paper.

The mathematics of renormalization is often motivated in language that is reminiscent of that of signal processing;
for a recent example among many,
\begin{quote}``the natural description of physics generally changes with the scale at which observations are made.
    Crudely speaking, this is no more high-minded a statement than saying that the world around us looks
    rather different when viewed through a microscope.
    More precisely, our parametrization of some system in terms of both the degrees of freedom and an action
    specifying how they interact generally change with scale.''~\cite[p. 178]{Rosten}
\end{quote}
Such a discussion may be subsumed by detailed engineering models that introduce different choices of window
functions both for measurements and for preparations to model the responses of different measurement apparatuses to
different preparation apparatuses.
The suggestion here ---unelaborated because, although this and the previous section are partly intended to establish
some connection with familiar topics, the intended main subject of this paper is contained in Section \ref{CAalgebra}---
is that in principle we would work only with window functions that we think appropriate to model an experiment, avoiding
the \emph{ad hoc} vagueness of a renormalization scale as a single parameter that determines broad aspects of the model;
indeed, sloganizing, if we use a different microscope we use a different window function.
If we want to use a renormalization scale in a pragmatic model, we might reasonably take it to be determined by
the maximum of the various characteristic frequencies of the test functions that would be used in a detailed model,
so that specifying a renormalization scale as well as a detailed model of an experiment overdetermines the
model unless the renormalization scale is close to the maximum frequency scales that are already present in the
detailed model.

\NewSection
\section{Nonlinear creation and annihilation $*$-algebras}\label{CAalgebra}
\renewcommand{\baselinestretch}{1.2}\normalsize
Insofar as we accept that interacting quantum fields are nonlinear functionals of test functions, we can introduce rather different
nonlinear models, grounded in the axioms of Wightman fields instead of in the renormalization group, which we pursue in the hope
that they might lead to otherwise inaccessible insight.
We begin here by noting that the conventional GNS-construction of a Fock space that is based on a
$*$-algebra of creation operators $a_{f_i}^\dagger$ and annihilation operators $a_{f_i}$, for some
countable set of test functions $\{f_i\}$, depends only on the commutator matrix
$[a^{\ }_{f_i},a_{f_j}^\dagger]=(f_i,f_j)$ being a positive semi-definite matrix, so that
$\VEV{a^{\ }_{f_1}\cdots a^{\ }_{f_m}a^\dagger_{g_1}\cdots a^\dagger_{g_n}}=\delta_{m,n}\mathsf{Per}\left[(f_i,g_j)\right]$,
the permanent of the matrix $(f_i,g_j)$, is a state over the algebra generated by the creation and annihilation $*$-algebra.
For the free field, this is ensured by $(f_i,f_j)$ being a Gram matrix of a positive semi-definite inner product
$(f,g)$ on the test function space that is diagonal in the wave-number basis,
$[a(k),a(k')^\dagger]=(2\pi)^4\delta^4(k-k')\itD_+(k)=(2\pi)^4\delta^4(k-k')2\pi\delta(k\!\cdot\!k-m^2)\theta(k_0)$; the factor
$\theta(k_0)$ implements the spectrum condition.
We also require, for locality for a quantum field $\xxi_f=a_{f^*}+a_f^\dagger$, that
$[\xxi_f,\xxi_g]=[a^{\ }_{f^*},a_g^\dagger]-[a^{\ }_{g^*},a_f^\dagger]$ is zero when $f$ and $g$ have space-like separated
supports, which for the free field is satisfied by $[a^{\ }_{f^*},a_g^\dagger]-[a^{\ }_{g^*},a_f^\dagger]=(f^*,g)-(g^*,f)$.
The notation $[a^{\ }_f,a_g^\dagger]=(f,g)$ refers to either a scalar or a non-scalar field, as discussed in
Appendix \ref{nonScalar}, but we will work with scalar fields, with the extension to non-scalar fields being
partly straightforward.
Extending this construction to the nonlinear case, we may also construct the commutator matrix
$[a^{\ }_{f_i},a_{f_j}^\dagger]$ as a positive semi-definite matrix by introducing Hadamard products such as
\begin{equation}\label{HadamardExample}
  [a^{\ }_{f_i},a_{f_j}^\dagger]=(f_i,f_j)_1+(f_i,f_j)_2(f_i,f_j)_3+(f_i,f_j)_4(f_i,f_j)_5+...+(f_i,f_j)_6(f_i,f_j)_7(f_i,f_j)_8+...,
\end{equation}
for some set of positive semi-definite inner products $(f_i,f_j)_n$ that satisfy the spectrum condition,
because, crucially for this construction, the Hadamard product $(M\circ N)_{ij}=M_{ij}N_{ij}$ of positive semi-definite
matrices is positive semi-definite~\cite[p. 141]{Bapat}.
We cannot use a matrix product, $\sum_k M_{ik}N_{kj}$, in this context because $[a^{\ }_{f_i},a^\dagger_{f_j}]$ must be
determined by just the two functions $f_i$ and $f_j$.

This construction straightforwardly satisfies locality because for each factor $(f,g)_n$, $(f^*,g)_n=(g^*,f)_n$ whenever
$f$ and $g$ have space-like separated supports.
For the spectrum condition, in preference to constructing an elaborate notation we rehearse the derivation
of the expected value of the 4-momentum for the lowest order non-vacuum states, using a normalized Hilbert space vector
$\hat\phi_{g_x}\Vacuum$, where we define translation of a test function implicitly in terms of the Fourier
transform as $\widetilde{g_x}(k)=\tilde g(k)\mathrm{e}^{-\mathrm{i}k\!\cdot\!x}$.
For the linear Wightman scalar field, for which $\VEV{\hat\phi^\dagger_f\hat\phi_g}=(f,g)$, we obtain for the 4-momentum
observable
\begin{eqnarray}\label{SpectrumEq}
  \VEV{\hat\phi^\dagger_{g_x}\mathrm{i}\frac{\partial\;}{\partial x^\mu}\hat\phi^{\ }_{g_x}}
         &=&\mathrm{i}\frac{\partial\;}{\partial x^\mu}\VEV{\hat\phi^\dagger_{g_y}\hat\phi^{\ }_{g_x}}\Big|_{y=x}
          = \mathrm{i}\frac{\partial\;}{\partial x^\mu}(g_y,g_x)\Big|_{y=x}\cr
         &=&\int k_\mu \tilde g^*(k)\itD_+(k)\tilde g(k)\frac{\Intd^4k}{(2\pi)^4},
\end{eqnarray}
which lies in or on the forward light-cone for all test functions $g$; the action of the 4-momentum operator as a derivation
ensures that the spectrum condition is satisfied for higher order states.
For nonlinear Wightman scalar fields, for example for the simplest case, $[a^{\ }_f,a^\dagger_g]=(f,g)^2$,
\begin{eqnarray}
   \mathrm{i}\frac{\partial\;}{\partial x^\mu}\VEV{\xxi^\dagger_{g_y}\xxi^{\ }_{g_x}}\Big|_{y=x}
         &=&\mathrm{i}\frac{\partial\;}{\partial x^\mu}(g_y,g_x)^2\Big|_{y=x}\cr
         &=&\int (k_1+k_2)_{\!\mu\;} \tilde g^*(k_1)\itD_+(k_1)\tilde g(k_1)\tilde g^*(k_2)\itD_+(k_2)\tilde g(k_2)
                         \frac{\Intd^4k_1}{(2\pi)^4}\frac{\Intd^4k_2}{(2\pi)^4},
\end{eqnarray}
which again lies in or on the forward light-cone for all test functions $g$; for higher order states, the 4-momentum operator
acts as a derivation on the $*$-algebra, as for the linear Wightman field case.
This direct approach to the 4-momentum operator in terms of translations makes no assumption that it can be defined as a
functional of the field.

A more interesting construction arises if we introduce a set of non-invariant individual positive semi-definite
inner products that are parameterized by wave-number,
\begin{equation}\label{IPu}
  (f,g)[u]=\int \tilde f^*(k)\itD_+(k-u)\tilde g(k)\frac{\Intd^4k}{(2\pi)^4}.
\end{equation}
We use these positive semi-definite inner products to construct, as the simplest ansatz,
\begin{eqnarray}\label{fgH}
  [a_f,a^\dagger_g]&=&(f,g)_H=\int (f,g)[u](f,g)[\lambda u]\tilde H(u)\frac{\Intd^4u}{(2\pi)^4}\cr
      &=&\int \tilde f^*(k_1)\itD_+(k_1-u)\tilde g(k_1)\tilde f^*(k_2)\itD_+(k_2-\lambda u)\tilde g(k_2)\tilde H(u)
              \frac{\Intd^4k_1}{(2\pi)^4}\frac{\Intd^4k_2}{(2\pi)^4}\frac{\Intd^4u}{(2\pi)^4}\cr
      &=&\int f^*(x_1)\iD_+(x_1-y_1)g(y_1)f^*(x_2)\iD_+(x_2-y_2)g(y_2)\cr
      &&\hspace{6em}\times\ H((x_1-y_1)+\lambda(x_2-y_2))
                          \Intd^4x_1\Intd^4y_1\Intd^4x_2\Intd^4y_2,\qquad
\end{eqnarray}
which is positive semi-definite and Lorentz invariant if $\tilde H(u)$ is.
We will find that we require $\lambda=-1$ to ensure that the spectrum condition is satisfied and $H(x)=H(-x)$
to ensure that the locality condition is satisfied.
When $f$ and $g$ have space-like separated supports, $(x_1-y_1)$ and $(x_2-y_2)$ are always both space-like in Eq. (\ref{fgH}),
in which case $\iD_+(x_1-y_1)=\iD_+(y_1-x_1)$ and $\iD_+(x_2-y_2)=\iD_+(y_2-x_2)$, so that
$[\xxi_f,\xxi_g]=[a_{f^*},a_g^\dagger]-[a_{g^*},a_f^\dagger]$ is zero when $f$ and $g$ have space-like separated
supports \emph{if} $H(x)=H(-x)$ for all $x$, because $(x_1-y_1)+\lambda(x_2-y_2)$ may be either space-like, light-like, or
time-like when $(x_1-y_1)$ and $(x_2-y_2)$ are space-like.
$H(x)$ in this case is not a distribution in real-space; see \cite{Wurm}, for example, from which we may also show that
because $\tilde H(u)\ge 0$, $H(x)$ at small separation is a positive multiple of $-(x^2)^{-1}$, for both massive and massless
cases, being always positive for small space-like separation and always negative for small time-like separation.
The introduction of negative frequency or space-like wave-number components, however, as well as being compatible
with locality, is also compatible with the spectrum condition because, repeating the process
introduced in Eq. (\ref{SpectrumEq}),
\begin{eqnarray}
         \mathrm{i}\frac{\partial\;}{\partial x^\mu}(g_y,g_x)_H\Big|_{y=x}
         &=&\int (k_1+k_2)_{\!\mu\;} \tilde g^*(k_1)\itD_+(k_1-u)\tilde g(k_1)\cr
         &&\hspace{4.5em}\times\  \tilde g^*(k_2)\itD_+(k_2-\lambda u)\tilde g(k_2)\tilde H(u)
                         \frac{\Intd^4k_1}{(2\pi)^4}\frac{\Intd^4k_2}{(2\pi)^4}\frac{\Intd^4u}{(2\pi)^4},
\end{eqnarray}
provided $\lambda=-1$, because in that case $k_1+k_2=k_1-u+k_2+u$ is in or on the forward light-cone because $k_1-u$
and $k_2+u$ are in or on the forward light-cone.
With this condition satisfied, despite the appearance of intermediate negative frequencies or space-like wave-number
components, the Hilbert space vectors nonetheless are always positive frequency.
The propagator $\tilde H(k)$ is effectively what may be called a ``hidden propagator'', insofar as it modifies the VEVs
of the nonlinear Wightman field by mediating a 4-momentum transfer without itself being associated with a field that is
measured.

We focus particular attention on the real-space expression for $(f,g)_H$,
\begin{eqnarray}\label{CleanAnsatz}
  (f,g)_H&=&\int f^*(x_1)\iD_+(x_1-y_1)g(y_1)f^*(x_2)\iD_+(x_2-y_2)g(y_2)\cr
      &&\hspace{6em}\times\ H((x_1-y_1)-(x_2-y_2))
                          \Intd^4x_1\Intd^4y_1\Intd^4x_2\Intd^4y_2,
\end{eqnarray}
noting that the hidden propagator factor can be rewritten as $H((x_1-x_2)-(y_1-y_2))$.
When the supports of $f$ and $g$ are at large space-like separation (when the VEVs of the linear Wightman field decrease
faster than exponentially with increasing separation) we can nonetheless engineer VEVs of the nonlinear Wightman field that are
very large by ensuring that terms for which $x_1-x_2$ is very close to $y_1-y_2$ dominate $(f,g)_H$, which we may
understand as analogous to a resonance condition between dipoles.
The VEVs nonetheless satisfy cluster decomposition because for any given local structure for $f$ and $g$,
increasing separation will arbitrarily minimize all VEVs.
With this, we can check off all the conditions that are required to allow the use of the Wightman reconstruction
theorem ---Cluster Decomposition, Relativistic Transformation, Spectrum, Hermiticity, Local Commutativity, and Positive
Definiteness---, excepting only, of course, that the VEVs are not distributions and we take these conditions in their
obvious smeared forms.
In their smeared forms, there is enough structure to allow the GNS-construction of a Hilbert space that generates a Gaussian
probability density for every operator $\xxi_f$.

The construction just given can be generalized to introduce arbitrary numbers of products of positive semi-definite
inner products $(f,g)_i[v_i]$, where the $v_i$ are arbitrary linear combinations of 4-momenta $u_j$ that
are contained in hidden propagator factors $\tilde H_j(u_j)$.
The undisplaced positive semi-definite inner products $(f,g)_i$ may correspond to different masses, with positive frequency
propagators $F_i(x-y)$, so that we obtain an expression
\begin{eqnarray}\label{GenIP1}
  (f,g)_{\{H\}}&=&\int\left[\prod\limits_{i=1}^n(f,g)_i{\scriptstyle\left[\sum\limits_{j=1}^m A_{ij}u_j\right]}\right]
                      \left[\prod\limits_{j=1}^m\tilde H_j(u_j)\frac{\Intd^4u_j}{(2\pi)^4}\right]\cr
               &=&\int\left[\prod\limits_{i=1}^n\tilde f^*(k_i)
                                                \tilde F_i\Bigl(k_i-{\scriptstyle\sum\limits_{j=1}^m A_{ij}u_j}\Bigr)
                                                \tilde g(k_i)\frac{\Intd^4k_i}{(2\pi)^4}\right]
                      \left[\prod\limits_{j=1}^m\tilde H_j(u_j)\frac{\Intd^4u_j}{(2\pi)^4}\right]\cr
               &=&\int\left[\prod\limits_{i=1}^n f^*(x_i)F_i(x_i-y_i)g(y_i)\Intd^4x_i\Intd^4y_i\right]
                      \left[\prod\limits_{j=1}^m H_j\Bigl({\scriptstyle\sum\limits_{i=1}^n A_{ij}(x_i-y_i)}\Bigr)\right].
\end{eqnarray}
$(f_i,f_j)_{\{H\}}$ is a positive semi-definite matrix because it is a positively weighted integral of Hadamard products of
positive semi-definite matrices; locality is satisfied if $H_j(-x)=H_j(x)$; and, to satisfy the spectrum condition, we
require that the column sums of the $A_{ij}$ are zero, $\sum\limits_{i=1}^n A_{ij}=0$, so that
\begin{equation}
  \sum\limits_{i=1}^n k_i=\sum\limits_{i=1}^n \left(k_i-{\scriptstyle\sum\limits_{j=1}^m A_{ij}u_j}\right),
\end{equation}
for all $u_j$, which was satisfied in the elementary example above, where the underlying integrand
was $(f,g)[u](f,g)[-u]$.

There is a possibility for extraordinary resonances between arbitrary powers of the test functions because of the
product $\prod\limits_{j=1}^m H_j\Bigl({\scriptstyle\sum\limits_{i=1}^n A_{ij}(x_i-y_i)}\Bigr)$, where the masses
and the pairwise relationships will determine a geometrical structure.
Given any theory we can add a resonance that can only be detected when the preparation and measurement test functions
satisfy very particular conditions, but that when those conditions are sufficiently satisfied the resonance will dominate
at arbitrary distances.
The range of possibilities is so large that it is clear that we must look for experimentally motivated
symmetries to bring some control and hopefully some tractability.
To construct a useful and verifiable theory we will have to take the structure of higher terms in the commutator to be
systematically generated, so that all the propagators $F_i(x-y)$ and $H_j(z)$ contained in $(f,g)_{\{H\}}$ are taken
from a small set and that the overall construction is invariant under appropriate symmetries, however model worlds are
possible in which there is no such systematization or in which the systematization is broken by spectacular terms.

The deformation of VEVs that is introduced by this construction includes nonlocal modifications of correlations
at space-like separation that are nonetheless local in the sense that no measurement incompatibility is introduced at
space-like separation, but as always for quantum mechanics it does not suggest or require a mechanism or an explanation for
the nonlocal correlations, it merely introduces models in which there are nonlocal correlations of the kind that this
particular mathematical structure allows, which we may then compare with such nonlocal correlations as appear in experiments.
As we saw in subsection \ref{DifferentQFT}, interacting Lagrangian QFT introduces a specific nonlocality additional to the
nonlocality already present in free linear Wightman fields, as it were by the introduction of free field operators everywhere
in the backward light-cone of an experiment, which modifies correlations without modifying measurement compatibility at
space-like separation, so an alternative mathematical structure should equally expect to modify correlations without
modifying measurement compatibility at space-like separation.

We are constructing functionals of two functions, subject to constraints that are only satisfied in the linear case by 
free quantum fields, but that are not especially hard to satisfy in the nonlinear case, so there is a superfluity of
possible constructions.
We mention here the lowest order example of a way to modify Eq. (\ref{IPu}) by the introduction of positive semi-definite
functions of $k\!\cdot\!u$, to give
\begin{equation}
  (f,g)''[u]=\int \tilde f^*(k)(k\!\cdot\!u)^2\tilde F(k-u)\tilde g(k)\frac{\Intd^4k}{(2\pi)^4},
\end{equation}
with obvious generalization of the quadratic form $(k\!\cdot\!u)^2$ to arbitrary positive polynomials,
which results in a greater contribution to $(f,g)''[u]$ by components for wave-numbers that are parallel to $u$.
$(f,g)''[u]$ is still diagonal in wave-number space and positive semi-definite, so it can be used to construct an analogue of
Eq. (\ref{fgH}),
\begin{eqnarray}
  &&\int (f,g)''[u](f,g)''[-u]\tilde H(u)\frac{\Intd^4u}{(2\pi)^4}\cr
      &&\quad=\int \tilde f^*(k_1)(k_1\!\cdot\!u)^2\tilde F(k_1-u)\tilde g(k_1)
              \tilde f^*(k_2)(k_2\!\cdot\!u)^2\tilde F(k_2+u)\tilde g(k_2)\tilde H(u)
              \frac{\Intd^4k_1}{(2\pi)^4}\frac{\Intd^4k_2}{(2\pi)^4}\frac{\Intd^4u}{(2\pi)^4}\cr
      &&\quad=\int f^*_{,\mu}(x_1)F(x_1-y_1)g^{\ }_{,\nu}(y_1)f^*_{,\alpha}(x_2)F(x_2-y_2)g^{\ }_{,\beta}(y_2)\cr
      &&\hspace{6em}\times\ H^{,\mu\nu\alpha\beta}((x_1-y_1)-(x_2-y_2))\Intd^4x_1\Intd^4y_1\Intd^4x_2\Intd^4y_2,
\end{eqnarray}
where $X^{\ }_{,\mu}$ is a tensor notation for a first-order derivative, \emph{etc}.
This construction modifies the resonance to be stronger when there are components of $f$ and $g$ at parallel wave-numbers.

Still further elaboration replaces the test functions by Lorentz invariant functionals of the test functions for which the
support is a subset of the causal completion of the original test function,
$\mathrm{Supp}(S[f])\subseteq\mathrm{Supp}(f)''$, and that ensure that the spectrum condition is satisfied, for 
which it is sufficient if $S_i[f]$ is translation covariant, $S_i[f_x]=S_i[f]_x$, to give
\begin{eqnarray}\label{GenIP2}
  (f,g)_{\{H,S\}}&=&\int\left[\prod\limits_{i=1}^n(S_i[f],S_i[g])_i{\scriptstyle\left[\sum\limits_{j=1}^m A_{ij}u_j\right]}\right]
                        \left[\prod\limits_{j=1}^m\tilde H_j(u_j)\frac{\Intd^4u_j}{(2\pi)^4}\right]\cr
                 &=&\int\left[\prod\limits_{i=1}^n\widetilde{S_i[f]}^*(k_i)
                                                  \tilde F_i\Bigl(k_i-{\scriptstyle\sum\limits_{j=1}^m A_{ij}u_j}\Bigr)
                                                  \widetilde{S_i[g]}(k_i)\frac{\Intd^4k_i}{(2\pi)^4}\right]
                        \left[\prod\limits_{j=1}^m\tilde H_j(u_j)\frac{\Intd^4u_j}{(2\pi)^4}\right]\cr
                 &=&\int\left[\prod\limits_{i=1}^n S_i[f]^*(x_i)F_i(x_i-y_i)S_i[g](y_i)\Intd^4x_i\Intd^4y_i\right]
                        \left[\prod\limits_{j=1}^m H_j\Bigl({\scriptstyle\sum\limits_{i=1}^n A_{ij}(x_i-y_i)}\Bigr)\!\right].
\end{eqnarray}
At simplest, the $S_i[f]$ may be arbitrary polynomials or analytic functions in $f$ and $f^*$ that preserve the vanishing of the
test function at infinity, but more elaborate constructions are possible, albeit they may result in even less tractability.

When we use interacting quantum fields as models, we typically introduce a list of wave-numbers $\{k_i\}$ and corresponding
widths as a shorthand for a list of test functions $\{f_i(k)\}$ that are more-or-less narrowly centered on those
wave-numbers and that are less-or-more spatially confined.
To model interactions in a nonlinear Wightman field formalism, we map these test functions into a space of operators,
which in the case just constructed may use Poincar\'e invariant scalar functionals $\lambda_i[f]$ and functionals $T_i[f]$
to give constructions such as, noting that normal-ordering may be defined exactly as for linear Wightman fields,
\begin{equation}\label{GenZeta}
  \zzeta_f=:\!\!\left[\lambda_0[f]\!+\!\lambda_1[f]\xxi_{T_1[f]}\!+\!\lambda_2[f]\xxi_{T_2[f]}\xxi_{T_3[f]}
                                                \!+\!\lambda_3[f]\xxi_{T_4[f]}\xxi_{T_5[f]}\!+\cdots
                                                +\!\lambda_4[f]\xxi_{T_6[f]}\xxi_{T_7[f]}\xxi_{T_8[f]}\!+\cdots\right]\!\!:\, ,
\end{equation}
which, we might say finally, results in non-Gaussian VEVs.
For the functionals $T_i[f]$, as for the $S_i[f]$ introduced in Eq. (\ref{GenIP2}), $T_i[f]$ must satisfy
$\mathrm{Supp}(T_i[f])\subseteq\mathrm{Supp}(f)''$ and must ensure that the spectrum condition is satisfied.
Transformations of the test function $f$ and of the functionals $T_i[f]$ can transform but cannot in general eliminate the
functionals $S_i[f]$ used in Eq. (\ref{GenIP2}).
The operators $\zzeta_f$ introduced by Eq. (\ref{GenZeta}) are qualitatively comparable to the interacting Lagrangian QFT
operators $\hat\xi_f$ constructed in Eq. (\ref{xiNO}), insofar as resonances between arbitrary numbers of points within the
supports of the test functions may be mediated by the various propagators and hidden propagators; there is a difference
that $\hat\xi_f$ is constructed using $\hat\phi(x)$ at every point in $\mathrm{Supp}(f\star G_{\mathrm{ret}})$ instead of
only at every point of $\mathrm{Supp}(f)''$, however renormalization introduces enough complications and $\zzeta_f$ works
in a different enough way that empirical comparison is nontrivial, and both complexes of operators satisfy the empirically
significant constraint of locality despite the difference.

We may also introduce a 4-momentum transfer between components of Eq. (\ref{GenZeta}) that can be compared with the
construction of Eq. (\ref{fgH}).
As a simplest example, using a free linear Wightman field $\hat\phi_f=a_{f^*}+a_f^\dagger$, $[a_f,a^\dagger_g]=(f,g)$,
we construct
\begin{equation}\label{z2def}
  \zzeta_f=\int\! :\!\hat\phi_{f_{[u]}}\hat\phi_{f_{[-u]}}\!:\tilde\mathsf{W}(u)\frac{\Intd^4u}{(2\pi)^4}
             =\int\! :\!\hat\phi(x_1)\hat\phi(x_2)\!:\!f(x_1)f(x_2)\mathsf{W}(x_1-x_2)\Intd^4x_1\Intd^4x_2,
\end{equation}
where $f_{[u]}(z)=f(z)\mathrm{e}^{\mathrm{i}u\cdot z}$, $\widetilde{f_{[u]}}(k)=\tilde f(k+u)$
{\small[this notation can be used to make the notation $(f,g)[u]$ redundant, if desired, because $(f,g)[u]=(f_{[u]},g_{[u]})$]}.
We note first that $\tilde\mathsf{W}(u)$ should be Lorentz invariant, but does not have to be positive semi-definite, so
that we will call these objects weights rather than propagators (although positive semi-definiteness in wave-number space is
not necessary for classical propagators, positive semi-definiteness is necessary where propagators are used to construct
a commutator between creation and annihilation operators), so that the behavior of $\mathsf{W}(x)$ at small space-like or
time-like separation is unconstrained, except that because of the symmetry of Eq. (\ref{z2def}), we can take
$\tilde\mathsf{W}(-u)=\tilde\mathsf{W}(u)$, $\mathsf{W}(-x)=\mathsf{W}(x)$, and hence also that
$\tilde\mathsf{W}^*(u)=\tilde\mathsf{W}(u)$ and $\mathsf{W}^*(x)=\mathsf{W}(x)$.
For the 2-measurement VEV $\VEV{\zzeta_f^\dagger\zzeta^{\ }_g}$, we have, in wave-number space and in real-space,
\begin{eqnarray}
  \VEV{\zzeta_f^\dagger\zzeta^{\ }_g}
      &=&\int\Bigl[(f_{[u_1]},g_{[u_2]})(f_{[-u_1]},g_{[-u_2]})+(f_{[u_1]},g_{[-u_2]})(f_{[-u_1]},g_{[u_2]})\Bigr]
             \tilde\mathsf{W}(u_1)\tilde\mathsf{W}(u_2)\frac{\Intd^4u_1}{(2\pi)^4}\frac{\Intd^4u_2}{(2\pi)^4}\cr
      &=&\int\left[\begin{array}{c}
                      \ \tilde f^*(k_1+u_1)\tilde F(k_1)\tilde g(k_1+u_2)\tilde f^*(k_2-u_1)\tilde F(k_2)\tilde g(k_2-u_2)\\
                       +\tilde f^*(k_1+u_1)\tilde F(k_1)\tilde g(k_1-u_2)\tilde f^*(k_2-u_1)\tilde F(k_2)\tilde g(k_2+u_2)
                   \end{array}\right]\cr
      &&\hspace{5em}\times\ \tilde\mathsf{W}(u_1)\tilde\mathsf{W}(u_2)
                    \frac{\Intd^4u_1}{(2\pi)^4}\frac{\Intd^4u_2}{(2\pi)^4}\frac{\Intd^4k_1}{(2\pi)^4}\frac{\Intd^4k_2}{(2\pi)^4}\cr
      &&\hspace{-6em}=2\!\int\! f^*(x_1)F(x_1-y_1)g(y_1)f^*(x_2)F(x_2-y_2)g(y_2)\mathsf{W}(x_1-x_2)\mathsf{W}(y_1-y_2)
              \Intd^4x_1\Intd^4y_1\Intd^4x_2\Intd^4y_2.
\end{eqnarray}
The free linear Wightman field in Eq. (\ref{z2def}) may be replaced by any of the nonlinear constructions we have already
introduced; if, for example, we replace the free Wightman field by $\xxi_f=a_{f^*}+a_f^\dagger$, $[a_f,a^\dagger_g]=(f,g)^2$,
we obtain for the 2-measurement VEV
\begin{eqnarray}
  \VEV{\!\zzeta_f^\dagger\zzeta^{\ }_g\!}
      &=&\!\!\!\int\!\Bigl[(f_{[u_1]},g_{[u_2]})^2(f_{[-u_1]},g_{[-u_2]})^2\!\!+\!(f_{[u_1]},g_{[-u_2]})^2(f_{[-u_1]},g_{[u_2]})^2\Bigr]
             \tilde\mathsf{W}(u_1)\tilde\mathsf{W}(u_2)\frac{\Intd^4u_1}{(2\pi)^4}\frac{\Intd^4u_2}{(2\pi)^4}\cr
      &&\hspace{-6em}=\!\!\int\!\left[\!\!\begin{array}{l}
      \hspace{0.75em}\tilde f^*(k_1+u_1)\tilde g(k_1+u_2)\tilde f^*(k_2-u_1)\tilde g(k_2-u_2)
                     \tilde f^*(k_3+u_1)\tilde g(k_3+u_2)\tilde f^*(k_4-u_1)\tilde g(k_4-u_2)\\
                    +\tilde f^*(k_1+u_1)\tilde g(k_1-u_2)\tilde f^*(k_2-u_1)\tilde g(k_2+u_2)
                     \tilde f^*(k_3+u_1)\tilde g(k_3-u_2)\tilde f^*(k_4-u_1)\tilde g(k_4+u_2)
                   \end{array}\!\!\right]\cr
      &&\hspace{3em}\times\ \tilde F(k_1)\tilde F(k_2)\tilde F(k_3)\tilde F(k_4)\tilde\mathsf{W}(u_1)\tilde\mathsf{W}(u_2)
                    \frac{\Intd^4k_1}{(2\pi)^4}\frac{\Intd^4k_2}{(2\pi)^4}\frac{\Intd^4k_3}{(2\pi)^4}\frac{\Intd^4k_4}{(2\pi)^4}
                    \frac{\Intd^4u_1}{(2\pi)^4}\frac{\Intd^4u_2}{(2\pi)^4}\cr
      &=&\!\int \left[\prod\limits_{i=1}^4f^*(x_i)F(x_i-y_i)g(y_i)\Intd^4x_i\Intd^4y_i\right]\cr
      &&\hspace{3em}\times\ \mathsf{W}(x_1-x_2+x_3-x_4)\Bigl[\mathsf{W}(y_1-y_2+y_3-y_4)+\mathsf{W}(y_1-y_2+y_4-y_3)\Bigr].
\end{eqnarray}
Such VEVs always result in expressions in which $\mathsf{W}(\cdot)$ is dependent on separations between the
$x_i$ integration variables and the $y_i$ integration variables separately, associated with different test functions separately
(because, in this example, the integration variable $u_1$ is a parameter only of $\tilde f$ and of $\tilde\mathsf{W}$, and similarly
$u_2$ is a parameter only of $\tilde g$ and of an independent instance of $\tilde\mathsf{W}$) instead of being dependent on
differences of separations associated with different test functions, such as the expression $(x_1-x_2)-(y_1-y_2)$ that occurred in
Eq. (\ref{CleanAnsatz}).
This construction, however, is not necessarily as trivial as the equivalent construction for the linear Wightman field, where
the result is just a function of the Gaussian observable, insofar as we take the Hilbert space of the theory to be generated
by the operators $\zzeta_f$, which in general is a subspace of the Hilbert space generated by Gaussian operators $\xxi_f$.

One consequence of nonlinearity is that a Hilbert space that supports a representation of the Poincar\'e group may
not be separable (that is, there may not be a countable basis), insofar as every multiple and sum of test functions
may construct an additional linearly independent operator.
Allowing non-separable Hilbert spaces is possible, and in time mathematical control of non-separable Hilbert spaces
may improve, but it would be preferable to introduce sufficient symmetries to restore separability of the Hilbert space.
It is also possible at the practical level to require only that the construction of the Hilbert space of a quantum field
theory is manifestly Lorentz and translation covariant, which is mathematically much weaker than requiring that the
Hilbert space support a representation of the Poincar\'e group.
Requiring only covariance is compatible with model-building in physics, which does not require that every possible
model that is related to a given model by Lorentz transformations or by translations is encompassed by a single
Hilbert space, only that a new Hilbert space can be readily constructed for any given Lorentz transformation or
translation.
Nonetheless, the introduction of symmetries that return us to the case of a separable Hilbert space that supports
a representation of the Poincar\'e group is desirable, for the sake both of mathematics and of engineering.

It is worth showing that thermal states, written formally as
$\omega_\beta(\hat A)=\frac{\mathsf{Tr}\left[\hat A\mathrm{e}^{-\beta\hat H}\right]}
                           {\mathsf{Tr}\left[\mathrm{e}^{-\beta\hat H}\right]}$, are Gaussian for a nonlinear Wightman field
$\xxi_f=a^{\ }_{f^*}+a^\dagger_f$ that has a nonlinear commutator $[a^{\ }_f,a^\dagger_g]=\nlIP{f,g}$.
[We note before proceeding that the Hamiltonian acts as an infinitesimal time-like translation, so that we have the algebraic
properties $\mathrm{e}^{-\beta\hat H}a^\dagger_f=a^\dagger_{\bbeta(f)}\mathrm{e}^{-\beta\hat H}$ and
$a_f\mathrm{e}^{-\beta\hat H}=\mathrm{e}^{-\beta\hat H}a_{\bbeta(f)}$, where $\Lbbeta(f)$
acts as a finite imaginary translation on $f$, so that in wave-number space, for a unit length time-like 4-vector $T$,
$\widetilde{\Lbbeta(f)}(k)=\mathrm{e}^{-\beta k\cdot T}\tilde f(k)$.]
The characteristic function for an observable $\xxi_f$ is
$\frac{\mathsf{Tr}\left[\mathrm{e}^{\mathrm{i}\lambda\hat\subxxi_f}\mathrm{e}^{-\beta\hat H}\right]}
      {\mathsf{Tr}\left[\mathrm{e}^{-\beta\hat H}\right]}$, for the numerator of which, taking both the cyclic property of
the trace and Baker-Campbell-Hausdorff identities to be formally applicable, we have (where an $\underbrace{\mbox{underbrace}}$
indicates the pair of operators that will be ``reversed''; where there is no underbrace the cyclic property of the trace will be used)
\begin{eqnarray}
  \mathsf{Tr}\left[\mathrm{e}^{\mathrm{i}\lambda\hat\subxxi_f}\mathrm{e}^{-\beta\hat H}\right]
      &=&    \mathsf{Tr}\left[\mathrm{e}^{\mathrm{i}\lambda a^\dagger_f}\mathrm{e}^{\mathrm{i}\lambda a_{f^*}}
                              \mathrm{e}^{-\beta\hat H}\right]\mathrm{e}^{-\lambda^2\nlIP{f^*,f}/2}\cr
      &=&    \mathsf{Tr}\left[\underbrace{\mathrm{e}^{\mathrm{i}\lambda a_{f^*}}\mathrm{e}^{-\beta\hat H}}
                              \mathrm{e}^{\mathrm{i}\lambda a^\dagger_f}\right]\mathrm{e}^{-\lambda^2\nlIP{f^*,f}/2}\cr
      &=&    \mathsf{Tr}\left[\mathrm{e}^{-\beta\hat H}\underbrace{\mathrm{e}^{\mathrm{i}\lambda a_{\sbbeta(f^*)}}
                              \mathrm{e}^{\mathrm{i}\lambda a^\dagger_f}}\right]\mathrm{e}^{-\lambda^2\nlIP{f^*,f}/2}\cr
      &=&    \mathsf{Tr}\left[\mathrm{e}^{-\beta\hat H}\mathrm{e}^{\mathrm{i}\lambda a^\dagger_f}
                              \mathrm{e}^{\mathrm{i}\lambda a_{\sbbeta(f^*)}}\right]
                        \mathrm{e}^{-\lambda^2\nlIP{f^*,f}/2-\lambda^2\nlIP{\bbeta(f^*),f}}\cr
      &=&    \mathsf{Tr}\left[\mathrm{e}^{\mathrm{i}\lambda a_{\sbbeta(f^*)}}\underbrace{\mathrm{e}^{-\beta\hat H}
                              \mathrm{e}^{\mathrm{i}\lambda a^\dagger_f}}\right]
                        \mathrm{e}^{-\lambda^2\nlIP{f^*,f}/2-\lambda^2\nlIP{\bbeta(f^*),f}}\cr
      &=&    \mathsf{Tr}\left[\underbrace{\mathrm{e}^{\mathrm{i}\lambda a_{\sbbeta(f^*)}}
                              \mathrm{e}^{\mathrm{i}\lambda a^\dagger_{\sbbeta(f)}}}\mathrm{e}^{-\beta\hat H}\right]
                        \mathrm{e}^{-\lambda^2\nlIP{f^*,f}/2-\lambda^2\nlIP{\bbeta(f^*),f}}\cr
      &=&    \mathsf{Tr}\left[\mathrm{e}^{\mathrm{i}\lambda a^\dagger_{\sbbeta(f)}}
                              \mathrm{e}^{\mathrm{i}\lambda a_{\sbbeta(f^*)}}\mathrm{e}^{-\beta\hat H}\right]
                        \mathrm{e}^{-\lambda^2\nlIP{f^*,f}/2-\lambda^2\nlIP{\bbeta(f^*),f}-\lambda^2\nlIP{\bbeta(f^*),\bbeta(f)}}\cr
      &&     \qquad\mbox{[repeating the above sequence, ...]}\cr
      &&\hspace{-8em}
            =\mathsf{Tr}\left[\mathrm{e}^{-\beta\hat H}\right]
                   \exp{\!\left(\!-\SmallFrac{1}{2}\lambda^2\nlIP{f^*,f}
                              -\lambda^2\sum\limits_{n=0}^\infty\left(\nlIP{\Lbbeta^{n+1}(f^*),\Lbbeta^n(f)}+
                                                                      \nlIP{\Lbbeta^{n+1}(f^*),\Lbbeta^{n+1}(f)}\right)\!\right)},
\end{eqnarray}
where we have written repeated application as $\widetilde{\Lbbeta^n(f)}(k)=\mathrm{e}^{-n\beta k\cdot T}\tilde f(k)$, which for
large $n$ approaches zero for all $k$ that are in or on the forward light-cone (and we take it that the test function space
requires that there is no zero wave-number component).
In the elementary case, test functions may have backward light-cone or space-like components, but those components do not
contribute to the physical Hilbert space because they are in the kernel of all inner products used in the construction of
nonlinear Wightman field commutators.
{\small[Eq. (\ref{GenIP1}) and similar constructions deserve special attention because space-like or backward
light-cone components of test functions make contributions to $(f,g)_{\{H\}}$, however in aggregate we have
\begin{eqnarray}
  (f,\Lbbeta^n(g))_{\{H\}}&=&\int\left[\prod\limits_{i=1}^n\tilde f^*(k_i)
                                                \tilde F_i\Bigl(k_i-{\scriptstyle\sum\limits_{j=1}^m A_{ij}u_j}\Bigr)
                                                \mathrm{e}^{-n\beta k_i\cdot T}\tilde g(k_i)\frac{\Intd^4k_i}{(2\pi)^4}\right]
                      \left[\prod\limits_{j=1}^m\tilde H_j(u_j)\frac{\Intd^4u_j}{(2\pi)^4}\right]\cr
               &=&\int\mathrm{e}^{-n\beta\sum_jk_j\cdot T}\left[\prod\limits_{i=1}^n\tilde f^*(k_i)
                                                \tilde F_i\Bigl(k_i-{\scriptstyle\sum\limits_{j=1}^m A_{ij}u_j}\Bigr)
                                                \tilde g(k_i)\frac{\Intd^4k_i}{(2\pi)^4}\right]
                      \left[\prod\limits_{j=1}^m\tilde H_j(u_j)\frac{\Intd^4u_j}{(2\pi)^4}\right],
\end{eqnarray}
where to satisfy the spectrum condition $\sum_jk_j$ is constructed to be in or on the forward light-cone, so that
$(f,\Lbbeta^n(g))_{\{H\}}$ approaches zero for large $n$; a similar discussion can be given for any nonlinear Wightman field,
because we always require the spectrum condition to be satisfied.]}
Translation invariance applied to the case of imaginary translations requires $\nlIP{\bbeta(f),g}=\nlIP{f,\bbeta(g)}$, so the
characteristic function for $\xxi_f$ in a thermal state is the Gaussian expression
\begin{equation}
  \frac{\mathsf{Tr}\left[\mathrm{e}^{\mathrm{i}\lambda\hat\subxxi_f}\mathrm{e}^{-\beta\hat H}\right]}
       {\mathsf{Tr}\left[\mathrm{e}^{-\beta\hat H}\right]}=\mathrm{e}^{-\Half\lambda^2\nlIP{f^*,f}_\beta},\qquad
                   \nlIP{f,g}_\beta\,\eqdef\,\nlIP{f,g}+2\sum\limits_{n=1}^\infty\nlIP{f,\Lbbeta^n(g)}.
\end{equation}
For the linear Wightman field case we obtain
\begin{equation}
  (f,g)_\beta=\int\tilde f^*(k)\itD_+(k)\coth(\SmallFrac{1}{2}\beta k\!\cdot\!T)\tilde g(k)\frac{\Intd^4k}{(2\pi)^4},
\end{equation}
however there is no comparably simple expression for the general nonlinear Wightman field case beyond the simple fact of
Gaussianity.
Nonetheless, we have the general property that, as for the linear Wightman field case, the thermal enhancement of
quantum fluctuations, determined by the ratio of the variances, $\nlIP{f^*,f}_\beta/\nlIP{f^*,f}$, is large for
small $\beta$ (high temperature) and for test functions that are dominated by their low frequency components.
The extension of the characteristic function for $\xxi_f$ to the multiple measurement case,
\begin{equation}\label{ThermalStateDef}
  \frac{\mathsf{Tr}\left[\mathrm{e}^{\mathrm{i}\sum_j\lambda_j\hat\subxxi_{f_j}}\mathrm{e}^{-\beta\hat H}\right]}
       {\mathsf{Tr}\left[\mathrm{e}^{-\beta\hat H}\right]}
           =\exp{\!\left(\!-\SmallFrac{1}{2}\sum\limits_{j,k}\lambda_j\lambda_k\nlIP{f_j^*,f_k}_\beta\!\right)},
\end{equation}
taken with the commutation relation $[\xxi_f,\xxi_g]=\nlIP{f^*,g}-\nlIP{g^*,f}$, is enough to fix the thermal state over
the algebra of observables generated by $\xxi_f$, and hence to fix the thermal state over the algebra of observables
generated by $\zzeta_f$.
We can contrast the thermal state with the vacuum state in more explicit terms by considering VEVs in terms of Gram matrix
permanents and the corresponding expressions for a Gaussian thermal state,
\begin{eqnarray}
  \VEV{a^\dagger_{g_i}\cdots a^\dagger_{g_n}a^{\ }_{f_1}\cdots a^{\ }_{f_m}}&=&0,\cr
  \VEV{a^{\ }_{f_1}\cdots a^{\ }_{f_m}a^\dagger_{g_i}\cdots a^\dagger_{g_n}}&=&\delta_{m,n}\mathsf{Per}\left[\nlIP{f_i,g_j}\right],\cr
  \mbox{so that }\VEV{\mathrm{e}^{\mathrm{i}\lambda a^{\ }_f}\mathrm{e}^{\mathrm{i}\lambda a^\dagger_g}}
                          &=&\sum\limits_{n=0}^\infty\frac{(-\lambda^2)^n\,n!\nlIP{f,g}^n}{n!^2}=\mathrm{e}^{-\lambda^2\nlIP{f,g}},\cr
  \omega_\beta(a^\dagger_{g_i}\cdots a^\dagger_{g_n}a^{\ }_{f_1}\cdots a^{\ }_{f_m})
              &=&\delta_{m,n}\mathsf{Per}\left[\SmallFrac{1}{2}\nlIP{f_i,g_j}_\beta-\SmallFrac{1}{2}\nlIP{f_i,g_j}_\beta\right],\cr
  \omega_\beta(a^{\ }_{f_1}\cdots a^{\ }_{f_m}a^\dagger_{g_i}\cdots a^\dagger_{g_n})
              &=&\delta_{m,n}\mathsf{Per}\left[\SmallFrac{1}{2}\nlIP{f_i,g_j}_\beta+\SmallFrac{1}{2}\nlIP{f_i,g_j}_\beta\right],\cr
  \mbox{so that }\omega_\beta(\mathrm{e}^{\mathrm{i}\lambda a^{\ }_f}\mathrm{e}^{\mathrm{i}\lambda a^\dagger_g})
                          &=&\mathrm{e}^{-\SmallFrac{1}{2}\lambda^2\nlIP{f,g}_\beta-\SmallFrac{1}{2}\lambda^2\nlIP{f,g}},
\end{eqnarray}
which, following the example of the KMS-condition~\cite[\S V.1]{Haag}, we may take to define the thermal state for nonlinear
Wightman fields instead of the relatively ill-defined Gibbs expression that we used formally to derive it.
Finally, we note that we can construct similar states for any operator $\hat X$ for which
$\mathrm{e}^{-\hat X}a^\dagger_f=a^\dagger_{X(f)}\mathrm{e}^{-\hat X}$ and
$a^{\ }_f\mathrm{e}^{-\hat X}=\mathrm{e}^{-\hat X}a^{\ }_{X(f)}$, where $X:f\mapsto X(f)$ satisfies the symmetry
$\nlIP{X(f),g}=\nlIP{f,X(g)}$, acts strongly enough as a contraction $\nlIP{f,X^n(g)}\rightarrow 0$ for the sum
$\sum_{n=1}^\infty\nlIP{f,X^n(g)}$ to converge, and ensures that $\nlIP{f_i,X^n(f_j)}$ is a positive semi-definite
matrix; in general, this construction will result in non-equilibrium thermodynamic states.
The amplitude transformation $X(f)=\mathrm{e}^{-\alpha}f$ is perhaps worth noting as enhancing quantum fluctuations
associated with linear terms in the commutator relative to the quantum fluctuations associated with higher degree
nonlinear terms.

A nonlinear Wightman field $\zzeta_w(x)=\zzeta_{w_x}$ does not in general satisfy an equation comparable to Eq. (\ref{XiEq}),
in which case we would have, for a functional $F_w$ and arbitrary test functions $f_i$,
\begin{equation}\label{XiEq}
  \VEV{\left[(\Box+m^2)\zzeta_w(x)+F_w\left(\zzeta_w(x),
           \frac{\partial\zzeta_w(x)}{\partial x_\mu}\frac{\partial\zzeta_w(x)}{\partial x^\mu},
           \frac{\partial^2\zzeta_w(x)}{\partial x_\mu\partial x_\nu}
                 \frac{\partial^2\zzeta_w(x)}{\partial x^\mu\partial x^\nu},...\right)\right]\zzeta_{f_1}\cdots \zzeta_{f_n}}=0,
\end{equation}
because $\zzeta_w(x)$ may be a functional of $\xxi_{T[w]_x}$ for translation invariant maps $T:w\mapsto T[w]$,
however the explicit constructions given above for $\zzeta_f$ fixes the VEVs of the field as a function of $x$ and as functionals
of $w$, so that the VEVs will satisfy equations of \emph{some} kind.
It presents a significant constraint on theories if we require the equations satisfied to be invariant for different choices of
the test functions $f_i$, or, less stringently, if we require the equations satisfied to be invariant for different choices of
$x_i$ when the $f_i$ are chosen to be $f_i=w_{x_i}$, for fixed $w$; although such requirements might be derived from experimental
data, it is not necessary from the perspective of the mathematical construction.

All the constructions in this section are by intention concrete, which is desirable to see the range of possible phenomenology
in moderate detail.
Taking Lorentz invariance to be more-or-less straightforwardly ensured by appropriate formalism and notation, the
conditions that a construction of a nonlinear Wightman field that uses creation and annihilation operators must satisfy
may be presented in summary form as
\begin{equation}
  \begin{array}{l}
    \xxi_f=a_{f^*}^{\ }+a_f^\dagger\\
    \left[a_f^{\ },a_g^\dagger\right]\!=\!\nlIP{f,g}\\
    a_f\left|0\right>=0
  \end{array}\quad
  \begin{array}{l l}
    \mbox{(CD)}&\bigl[a_f^{\ },a_g^\dagger\bigr]\rightarrow 0\mbox{ as space-like separation}\rightarrow\infty\\
    \mbox{(PI)}&\bigl[a_{f_x}^{\ },a_{g_x}^\dagger\bigr]=\bigl[a_f^{\ },a_g^\dagger\bigr]\\
    \mbox{(Sp)}&\VEV{\hat A_x^\dagger\mathrm{i}\partial_\mu\hat A_x}\mbox{ is in or on the forward light-cone}\\
    \mbox{(H)}&\bigl[a_f^{\ },a_g^\dagger\bigr]^*=\bigl[a_g^{\ },a_f^\dagger\bigr]\\
    \mbox{(Loc)}&\bigl[a_{f^*}^{\ },a_g^\dagger\bigr]=\bigl[a_{g^*}^{\ },a_f^\dagger\bigr]
                 \mbox{ if $f$ and $g$ are space-like separated}\\
    \mbox{(Pos)}&\bigl[a_{f_i}^{\ },a_{f_j}^\dagger\bigr]\mbox{ is a positive semi-definite matrix}
  \end{array}
\end{equation}
where the operator $\hat A_x$ is any operator constructed using test functions $f_i$ transformed to use the same test functions
displaced by $x$.
The observable field $\zzeta_f$ is a functional of Gaussian fields $\xxi_{T_i[f]}$ in any combination
that preserves Poincar\'e invariance and locality, $[\zzeta_f,\zzeta_g]=0$ if $f$ and $g$ are space-like separated.

\NewSection
\section{Discussion}
The next steps for this approach are to characterize as precisely as possible how close nonlinear Wightman fields ---which
clearly introduce significantly more flexibility than most axiomatic approaches--- can come to equalling the modeling
effectiveness of Lagrangian QFT and of statistical physics; and to investigate what natural constraints may be provided
by symmetry or other empirically motivated considerations.
It is to be hoped that the investigation will clarify our understanding of Lagrangian QFT even if we discover
that nonlinear Wightman fields are not a useful modeling tool for particle physics experiments.
It also appears that some concepts in statistical physics might be clarified by comparison with the introduction of
hidden propagators, which introduce potentially complex stochastic synchronizations that do not modify measurement
incompatibility at space-like separation but may modify observed correlations even at large space-like separations
when enough care is taken to construct resonances between preparation and measurement apparatuses.

What equivalences there might be between models is complicated by the possibility of making different assignments of frequencies,
widths, and other, more precise details when choosing what test functions to use when constructing a model for a given
experimental apparatus than we have been accustomed to making when constructing models in interacting Lagrangian QFT.
We not only have to consider whether there is a model in which $\zzeta_f\thickapprox\hat\xi_f$ for any test function $f$ and for
some empirically successful Lagrangian QFT operator $\hat\xi_f$, where $\thickapprox$ denotes some kind of empirical equivalence,
we also have to consider whether there is a consistent choice of operators $\hat A_\subzzeta$ and $\hat A_\xi$ for a given
preparation apparatus or measurement apparatus, what we will call ``engineering rules'' for the $*$-algebras generated by
$\zzeta_f$ and by $\hat\xi_f$, for which $\hat A_\subzzeta\thickapprox\hat A_\xi$; an equivalence $\zzeta_f\thickapprox\hat\xi_f$
is desirable but physically inessential, whereas for a theory to be useful there must be an empirical correspondence
$\hat A_\subzzeta\thickapprox\hat A_\xi$ for a substantial range of preparation and measurement apparatuses.

The primary connection with the past that is preserved here has been the Wightman axioms, with the introduction
of \emph{some} form of nonlinearity suggested by a specific reading of the mathematics of the renormalization of
Lagrangian QFT.
The secondary connection with the past that is preserved here has been the structure of a creation and annihilation
operator $*$-algebra, the only known model of the Wightman axioms in Minkowski space, modified by the introduction
of a natural nonlinearity.
The resulting mathematics is of some interest, however it is probably too unconstrained, needing either a further or a
different connection with existing theory and with experiment.
As far as introducing different choices is concerned, a reliance on creation and annihilation operators as a starting point
seems relatively unimaginative, given the variety of algebras that is available in the mathematical literature, although
they seem a good enough first choice, whereas gauge invariance is the obvious further choice (a first essay at which may
be found at the end of Appendix \ref{nonScalar}).

If we find that nonlinear Wightman fields \emph{are} useful, their considerable and perhaps excessive flexibility suggests
that they may be an elaborate intermediate system of \textsl{epicycles} more than an immediately explanatory theory;
some form of higher level mathematics will likely have to be introduced to provide an explanatory structure.
Nonlinear Wightman field models are apparently adaptable to a wide range of phenomena, comparably to the adaptability of
nonlinear differential equations as deterministic models.
The greater flexibility as a starting point for models is a reasonable reflection of the higher order of the
mathematics of probabilities over the configuration space of a field theory, and the non-classical description of the
evolution of those probability densities over time that is afforded by quantum field theory, compared to the lower order
mathematics of the classical dynamics of particles or of a field configuration and momentum.

I am grateful for comments from David Schneider on an earlier draft.

\appendix
\section{Non-scalar fields}\label{nonScalar}
The notation $[a^{\ }_f,a_g^\dagger]=(f,g)$ for the free field case, and equally for the nonlinear analogues in Section
\ref{CAalgebra}, applies straightforwardly to observable non-scalar fields by taking test functions effectively to be
functions of their Lorentz or internal symmetries as well as of position, so that a field operator such as $\hat\phi_f$
or an annihilation operator such as $a_f$ is a scalar object.
The free scalar field of a given mass is associated with the scalar positive semi-definite inner product
\begin{equation}
  [a^{\ }_f,a_g^\dagger]=(f,g)=\int\tilde f^*(k)\tilde g(k)2\pi\delta(k\!\cdot\!k-m^2)\theta(k_0)\frac{\Intd^4k}{(2\pi)^4};
\end{equation}
for the free electromagnetic or massive bivector field, this is replaced by the scalar positive semi-definite inner product
\begin{equation}\label{EMcomm}
  [a^{\ }_f,a_g^\dagger]=(f,g)=-\int\tilde f_{[\alpha\mu]}^*(k)k^\alpha g^{\mu\nu}k^\beta\tilde g_{[\beta\nu]}(k)
             2\pi\delta(k\!\cdot\!k-m^2)\theta(k_0)\frac{\Intd^4k}{(2\pi)^4},
\end{equation}
where the positive semi-definiteness arises from $\tilde f_{[\alpha\mu]}^*(k)k^\alpha$ and
$k^\beta\tilde g_{[\beta\nu]}(k)$ being space-like 4-vectors that are both orthogonal to the null or time-like 4-vector $k$
(note that the bivector test function $g_{[\beta\nu]}(\cdot)$ must be distinguished from the constant Minkowski space metric
$g^{\mu\nu}$, with signature $(+---)$, following the conventions in~\cite{IZ}).
Hence, the notation $[a^{\ }_f,a_g^\dagger]=(f,g)$ is equally applicable to the scalar, to the electromagnetic field, and
for any test function space for which we introduce a positive semi-definite inner product.
When hidden propagators $H(u)$ and wave-number shifted positive semi-definite inner products are introduced, a wave-number
index $u$ might be taken to include Lorentz or internal indices, provided positive semi-definiteness is maintained, however
the simplest construction introduces the wave-number shifted positive semi-definite inner product
\begin{equation}
  (f,g)[u]=-\int\tilde f_{[\alpha\mu]}^*(k)(k-u)^\alpha g^{\mu\nu}(k-u)^\beta\tilde g_{[\beta\nu]}(k)
             2\pi\delta((k-u)^2-m^2)\theta(k_0-u_0)\frac{\Intd^4k}{(2\pi)^4}
\end{equation}
for the electromagnetic field, indexed only by the wave-number $u$.

\newcommand\aOP{{\mathbf{a}}}
For charged fields, reproducing the conventional model in a more compact and more manageable notation, we introduce a
column vector $f=\left({f_1\atop f_2}\right)$, the Hermitian conjugate $f^\dagger=(f_1^*,f_2^*)$, and a charge
conjugation involution, $f^c=\left({f_2^*\atop f_1^*}\right)$, with the field defined as
\begin{equation}
  \hat\phi_f=\aOP^{\ }_{f^c}+\aOP^\dagger_f=a^{\ }_{1,f_2^*}+a^{\ }_{2,f_1^*}+a^\dagger_{1,f_1}+a^\dagger_{2,f_2}
\end{equation}
and with the commutation relations
\begin{equation}
  [\aOP^{\ }_f,\aOP^\dagger_g]=(f,g)=\int \tilde f^\dagger(k)\itD_+(k)\tilde g(k)\frac{\Intd^4k}{(2\pi)^4},
\end{equation}
so that $[\hat\phi_f,\hat\phi_g]=(f^c,g)-(g^c,f)=(f_2^*,g_1)-(g_1^*,f_2)+(f_1^*,g_2)-(g_2^*,f_1)$, which is zero, satisfying
locality, when the supports of $f$ and $g$ are space-like separated.
Additionally, note that $\hat\phi_f$ is an observable, $\hat\phi^{\ }_f=\hat\phi^\dagger_f$, when
$f=f^c=\left({f_1\atop f_1^*}\right)$.

\newcommand\bOP{{\mathbf{b}}}
\newcommand\dOP{{\mathbf{d}}}
We can also construct free Fermionic Dirac fields in a similar smeared operator form.
We take $\hat\psi_U$ to be a complex linear functional of a spinor-valued test function $U$,
$\hat\psi^{\ }_{U^c}=\bOP^{\ }_{U^c}+\dOP_U^\dagger$, where the annihilation operators $\bOP^{\ }_U$ and $\dOP^{\ }_U$
follow the usual convention of being complex anti-linear in $U$, so that we have to introduce charge conjugation to ensure
that $\bOP_{U^c}$ is complex linear in $U$.
For these scalar smeared operators, we have the anti-commutation relations
\begin{eqnarray}
  \{\bOP^{\ }_U,\bOP_V^\dagger\}=\{\dOP^{\ }_U,\dOP_V^\dagger\}
                                        &=&\int\frac{\Intd^4k}{(2\pi)^4}2\pi\delta(k\!\cdot\!k-m^2)
                              \theta(k_0)\overline{\tilde U(k)}(k_\mu\gamma^\mu+m)\tilde V(k)\ \eqdef\ (U,V)_+,\label{BDeq}\\
  \{\hat\psi^{\ }_U,\hat\psi_V^\dagger\}&=&\{\bOP^{\ }_{U^c},\bOP_{V^c}^\dagger\}+\{\dOP_U^\dagger,\dOP^{\ }_V\}\cr
                           &=&\int\frac{\Intd^4k}{(2\pi)^4}2\pi\delta(k\!\cdot\!k-m^2)
                              \varepsilon(k_0)\overline{\tilde V(k)}(k_\mu\gamma^\mu+m)\tilde U(k)\ \eqdef\ (V,U),
\end{eqnarray}
where $\{\bOP^{\ }_{U^c},\bOP_{V^c}^\dagger\}$ and $\{\dOP_U^\dagger,\dOP^{\ }_V\}$, which are both positive semi-definite,
contribute the positive and negative frequency parts of the positive semi-definite anti-commutator
$\{\hat\psi^{\ }_U,\hat\psi_V^\dagger\}$, which is zero when $U$ and $V$ have space-like separated supports.
The arbitrary phase introduced by charge conjugation cancels in these equations because of the identities
$\overline{A^c}\gamma^\mu B^c=\overline{B}\gamma^\mu A$ and $\overline{A^c}\,B^c=-\overline{B}\,A$.
Explicitly, for $\{\bOP_{U^c}^{\ },\bOP^\dagger_{V^c}\}$, we have
\begin{eqnarray}\label{UcVc}
  \{\bOP^{\ }_{U^c},\bOP^\dagger_{V^c}\}&=&(U^c,V^c)_+\cr
                                   &=&\int\frac{\Intd^4k}{(2\pi)^4}2\pi\delta(k\!\cdot\!k-m^2)
                                      \theta(k_0)\overline{\widetilde{U^c}(k)}(k_\mu\gamma^\mu+m)\widetilde{V^c}(k),\cr
                                   &=&\int\frac{\Intd^4k}{(2\pi)^4}2\pi\delta(k\!\cdot\!k-m^2)
                                      \theta(k_0)\overline{\left[\widetilde{V^c}(k)\right]^c}(k_\mu\gamma^\mu-m)
                                                           \left[\widetilde{U^c}(k)\right]^c,\cr
                                   &=&\int\frac{\Intd^4k}{(2\pi)^4}2\pi\delta(k\!\cdot\!k-m^2)
                                      \theta(k_0)\overline{\tilde V(-k)}(k_\mu\gamma^\mu-m)\tilde U(-k),\cr
                                   &=&\int\frac{\Intd^4k}{(2\pi)^4}2\pi\delta(k\!\cdot\!k-m^2)
                                      \theta(-k_0)\overline{\tilde V(k)}(-k_\mu\gamma^\mu-m)\tilde U(k)\ \eqdef\ (V,U)_-,
\end{eqnarray}
and $(V,U)=(V,U)_++(V,U)_-$.
It is essential for the construction of the Fermionic Fock space that Eq. (\ref{BDeq}) is positive semi-definite and that
when $U$ and $V$ have space-like separated supports we have the identity $(V,U)_+=-(V,U)_-$.
We can construct a generalization to nonlinear Wightman fields similar to that of Section \ref{CAalgebra} by
taking, for example, $\{\bOP^{\ }_U,\bOP_V^\dagger\}=\{\dOP^{\ }_U,\dOP_V^\dagger\}=[(U,V)_+]^3$, so that
$\{\hat\psi^{\ }_U,\hat\psi_V^\dagger\}=[(V,U)_+]^3+[(V,U)_-]^3$, which is zero when $U$ and $V$ have space-like
separated supports; it is apparent that only odd powers of $(U,V)_+$ may be introduced.
For even powers of $(U,V)_+$, we may construct charged Bosonic fields, such as, for example,
$[b^{\ }_U,b^\dagger_V]=[d^{\ }_U,d^\dagger_V]=[(U,V)_+]^2$, $\xxi^{\ }_U=b^{\ }_{U^c}+d^\dagger_U$, for
which $[\xxi^{\ }_U,\xxi^\dagger_V]=[(U^c,V^c)_+]^2-[(V,U)_+]^2=[(V,U)_-]^2-[(V,U)_+]^2$, which satisfies locality.

Eq. (\ref{BDeq}) is not the most general positive semi-definite inner product on the Dirac spinor test function space that
satisfies locality;
if we allow the use of $\gamma^5=\mathrm{i}\gamma^0\gamma^1\gamma^2\gamma^3$, for which $(\gamma^5)^2=1$,
$\overline{\gamma^5}=-\gamma^5$, $\overline{A^c}\gamma^\mu\gamma^5 B^c=\overline{B}\gamma^\mu\gamma^5 A$,
$\overline{A^c}\gamma^5 B^c=-\overline{B}\gamma^5 A$, and $\gamma^5\gamma^\mu+\gamma^\mu\gamma^5=0$, then we can
construct
\begin{equation}
  \int\!\frac{\Intd^4k}{(2\pi)^4}2\pi\delta(k\!\cdot\!k-m^2)\theta(k_0)\overline{\tilde U(k)}
        \left(k_\mu\gamma^\mu+(\alpha_{{}_1}m+\alpha_{{}_2}m\mathrm{i}\gamma^5+\alpha_{{}_3}k_\mu\gamma^\mu\gamma^5)\right)
        \tilde V(k)\ \eqdef\ (U,V)_{\underline{\alpha}+},
\end{equation}
which is positive semi-definite provided $\alpha_{{}_1}^2+\alpha_{{}_2}^2+\alpha_{{}_3}^2\le 1$, for which
\begin{eqnarray}
  \{\hat\psi^{\ }_U,\hat\psi^\dagger_V\}
              &=&\int\!\frac{\Intd^4k}{(2\pi)^4}2\pi\delta(k\!\cdot\!k-m^2)\varepsilon(k_0)\overline{\tilde U(k)}
        \left(k_\mu\gamma^\mu+(\alpha_{{}_1}m+\alpha_{{}_2}m\mathrm{i}\gamma^5+\alpha_{{}_3}k_\mu\gamma^\mu\gamma^5)\right)
                               \tilde V(k)\cr
              &\eqdef&(V,U)_{\underline{\alpha}}.
\end{eqnarray}
If we insist that this construction projects to a 2-dimensional subspace, then we must have
$\alpha_{{}_1}^2+\alpha_{{}_2}^2+\alpha_{{}_3}^2=1$, with the usual massive Dirac spinor field being the case
$(U,V)_+=(U,V)_{(1,0,0)+}$.
A non-zero value for $(\alpha_{{}_1},\alpha_{{}_2},\alpha_{{}_3})$ explicitly breaks the SU(2) symmetry group generated by
$\mathrm{i}\gamma^5$, $\frac{\mathrm{i} k_\mu\gamma^\mu}{m}$, and $\frac{k_\mu\gamma^\mu\gamma^5}{m}$, whereas
\begin{equation}
  \int\frac{\Intd^4k}{(2\pi)^4}2\pi\delta(k\!\cdot\!k-m^2)
                              \theta(k_0)\overline{\tilde U(k)}k_\mu\gamma^\mu\tilde V(k)\ \eqdef\ (U,V)_{\underline{0}+}
\end{equation}
is positive semi-definite and SU(2) invariant.
We might use this structure together with a hidden propagator to introduce a nonlinear Wightman field with, as an example
taken from a plethora of possibilities, a commutator that has a first-degree component with explicitly broken SU(2) symmetry
and a third-degree component that is SU(2) invariant,
\begin{equation}
  (U,V)_+\,+\!\!\int\!(U,V)_{\underline{\alpha}+}[u](U,V)_{\underline{\alpha}+}[-u]
         \delta(\alpha_{{}_1}^2+\alpha_{{}_2}^2+\alpha_{{}_3}^2-1)\Intd^3\alpha\tilde H(u)\frac{\Intd^4u}{(2\pi)^4}(U,V)_{\underline{0}+}.
\end{equation}

The highly developed computational technology associated with Feynman diagrams does not encourage the development of
formalisms that use test functions.
Test function or window function thinking is significantly different from quantum field thinking, however, leading, for example,
to a construction for gauge invariant nonlinear Wightman fields in which we introduce a Dirac spinor test function $U(x)$, a
scalar test function $u(x)$, and a 4-vector test function $\mathfrak{u}_\mu(x)$, with which we locally construct a Dirac spinor test
function $[\mathrm{e}^{\mathrm{i}u}U](x)=\mathrm{e}^{\mathrm{i}u(x)}U(x)$ and a 4-vector test function
$\left[\frac{\partial u}{\partial x^\mu}-\mathfrak{u}_\mu\right](x)$.
With the triplet $(U,u,\mathfrak{u})$ and a similar triplet $(V,v,\mathfrak{v})$, we can replace Eq. (\ref{BDeq}) by a
locally gauge invariant set of two anticommutators and a commutator,
\begin{eqnarray}
  \{\bOP^{\ }_{(U,u)},\bOP_{(V,v)}^\dagger\}&=&\{\dOP^{\ }_{(U,u)},\dOP_{(V,v)}^\dagger\}=((U,u),(V,v))_+
                 \,\eqdef\,(\mathrm{e}^{\mathrm{i}u}U,\mathrm{e}^{\mathrm{i}v}V)_+\cr
  &&\hspace{-4em}=\int\frac{\Intd^4k}{(2\pi)^4}2\pi\delta(k\!\cdot\!k-m^2)
                              \theta(k_0)\overline{\widetilde{[\mathrm{e}^{\mathrm{i}u}U]}(k)}(k_\mu\gamma^\mu+m)
                                                   \widetilde{[\mathrm{e}^{\mathrm{i}v}V]}(k),\cr
  [a^{\ }_{(u,\mathfrak{u})},a^\dagger_{(v,\mathfrak{v})}]&=&((u,\mathfrak{u}),(v,\mathfrak{v}))
                \,\eqdef\,(\partial_\mu u\!-\!\mathfrak{u}_\mu,\partial_\nu v\!-\!\mathfrak{v}_\nu)\cr
  &&\hspace{-4em}=\int\frac{\Intd^4k}{(2\pi)^4}2\pi\delta(k\!\cdot\!k)\theta(k_0)
                                           \widetilde{[\partial_\mu u\!-\!\mathfrak{u}_\mu]}{}^*(k)
                                           k^\mu k^\nu\widetilde{[\partial_\nu v\!-\!\mathfrak{v}_\nu]}(k),
\end{eqnarray}
to which must be added nonlinear interaction terms, potentially including hidden propagators, all of which must be invariant
under the local gauge symmetry
\begin{equation}
  U(x)\rightarrow \mathrm{e}^{\mathrm{i}\phi(x)}U(x),
     \qquad u(x)\rightarrow u(x)-\phi(x),
     \qquad \mathfrak{u}_\mu(x)\rightarrow \mathfrak{u}_\mu(x)-\partial_\mu\phi(x),
     \qquad \phi(x)\in\mathbb{R},
\end{equation}
and similarly for $(V,v,\mathfrak{v})$.
We have delayed describing $\mathfrak{u}_\mu(x)$ and $\mathfrak{v}_\nu(x)$ until after this equation because the mathematics
makes it clear they are \emph{test connections}, objects that do not exist in Lagrangian QFT, which, like a test function,
can be succinctly said to be a description or coordinates of what we are measuring instead of what the field is.
Measurement operators can be constructed from these creation and annihilation operators,
\begin{eqnarray}
  \hat\psi_{(\underline{U},u,\mathfrak{u})}&=&\bOP^{\ }_{(U_1^c,-u^*)}+\dOP_{(U_1,u)}^\dagger
                                             +\dOP^{\ }_{(U_2^c,-u^*)}+\bOP_{(U_2,u)}^\dagger,\cr
  \xxi_{(\underline{U},u,\mathfrak{u})}&=&a^{\ }_{(-u^*,-\mathfrak{u}^*)}+a^\dagger_{(u,\mathfrak{u})},
\end{eqnarray}
where we have introduced a vector of two Dirac spinor test functions $\underline{U}=(U_1,U_2)$, because the Dirac
wave function is charged, together with just one scalar and 4-vector test function pair.
Note that $\hat\psi_{(\underline{U},u,\mathfrak{u})}$ is an observable,
$\hat\psi_{(\underline{U},u,\mathfrak{u})}^\dagger=\hat\psi^{\ }_{(\underline{U},u,\mathfrak{u})}$, just if
$[\mathrm{e}^{\mathrm{i}u(x)}U_1(x)]^c=\mathrm{e}^{\mathrm{i}u(x)}U_2(x)$, and that with the addition of nonlinear
interaction terms, $\hat\psi_{(\underline{U},u,\mathfrak{u})}$ and $\xxi_{(\underline{U},u,\mathfrak{u})}$ may both
depend on all their parameters.
Note also that the signs introduced into the two conjugations $(U^c,-u^*)$ and $(-u^*,-\mathfrak{u}^*)$ are required
to ensure that both observables are local.
It is sometimes claimed that local gauge symmetry is the mathematical structure that underlies the empirical success
of Lagrangian QFT, which is taken here to be enough to motivate further investigation of this or other locally gauge
invariant constructions in a test function framework, although it is far from clear how tractable or empirically successful
the mathematics will be.
Whether and how any of this multiplicity of structure might be fitted together to make contact with experiment and to give
a detailed alternative to quantum electrodynamics or to the standard model of particle physics is of course not yet
in reach.

\end{document}